\documentclass[manuscript,onecolumn]{acmart}

\usepackage{etoolbox}
\usepackage{placeins} 
\usepackage{soul}
\usepackage{ulem}
\AtBeginDocument{%
  \providecommand\BibTeX{{%
    \normalfont B\kern-0.5em{\scshape i\kern-0.25em b}\kern-0.8em\TeX}}}

\setcopyright{acmcopyright}
\copyrightyear{2018}
\acmYear{2018}
\acmDOI{XXXXXXX.XXXXXXX}

\acmJournal{POMACS}
\acmVolume{37}
\acmNumber{4}
\acmArticle{111}
\acmMonth{8}

\newcommand{\add}[1]{{#1}}
\newcommand{\anedit}[1]{{#1}}

\newcommand{\replace}[2]{#2}
\newcommand{\del}[1]{}

\begin{document}
\title{Supporting Gig Worker Needs and Advancing Policy Through Worker-Centered Data-Sharing}

\author{Jane Hsieh}
\authornote{Co-authors contributed equally to this research.}
\affiliation{%
  \institution{Carnegie Mellon University}
  \city{Pittsburgh, PA}
  \country{USA}
}

\author{Angie Zhang}
\authornotemark[1]
\affiliation{%
  \institution{University of Texas at Austin}
  \city{Austin, TX}
  \country{USA}
  }

\author{Mialy Rasetarinera}
\authornote{Most of this research was conducted while the author was affiliated with UC Berkeley.}
\affiliation{%
  \institution{University of Pennsylvania}
  \country{USA}
  \city{Berkeley, CA}
  }

\author{Erik Chou}
\affiliation{%
  \institution{University of Maryland}
  \city{College Park, MD}
  \country{USA}
}

\author{Daniel Ngo}
\affiliation{%
  \institution{University of Minnesota}
  \city{Minneapolis, MN}
  \country{USA}
}

\author{Karen Lightman}
\affiliation{
\institution{Carnegie Mellon University}
\city{Pittsburgh, PA}
\country{USA}
}

\author{Min Kyung Lee}
\affiliation{%
  \institution{University of Texas at Austin}
  \city{Austin, TX}
  \country{USA}}

\author{Haiyi Zhu}
\affiliation{%
  \institution{Carnegie Mellon University}
  \city{Pittsburgh, PA}
  \country{USA}
}

\renewcommand{\shortauthors}{Hsieh \& Zhang et al.}
\begin{abstract}
The proliferating adoption of platform-based gig work increasingly raises concerns for worker conditions. Past studies documented how platforms leveraged design to exploit labor,  withheld information to generate power asymmetries, and left workers alone to manage logistical overheads as well as social isolation. However, researchers also called attention to the potential of helping workers overcome such costs via worker-led datasharing, which can enable collective actions and mutual aid among workers, while offering advocates, lawmakers and regulatory bodies insights for improving work conditions. To understand stakeholders' desiderata for a data-sharing system (i.e. functionality and policy initiatives that it can serve), we interviewed 11 policy domain experts in the U.S. and conducted co-design workshops with 14 active gig workers across four domains. Our results outline policymakers’ prioritized initiatives, information needs, and (mis)alignments with workers’ concerns and desires around data collectives. We offer design recommendations for data-sharing systems that support worker needs while bringing us closer to legislation that promote more thriving and equitable gig work futures.
\end{abstract}

\maketitle

\section{Introduction}\label{intro}
Gig work, often characterized by its short-term and flexible tasks performed by independent contractors, has transformed the work practices of individuals across the globe. \replace{Over the past decade, g}{G}ig workers increasingly occupy and reshape the workforce---expanding work access to individuals at the margins while providing rapidly-supplied labor to consumers and businesses \cite{pZ2o,OBVk,EGjj,gkI5}. However, recent accounts show that through algorithmic management, gig platforms shift many costs and risks of employment onto workers alone, harming their health, safety\replace{ and}{,} financial security \replace{as well as}{and general} well-being\del{ in general} \cite{OBVk,gkI5,bSah,uoX5,8NZx,TdXv}.

In light of increasing concern for gig worker rights and conditions, scholars from various nations\del{(in e.g. Australia, Europe and North America} \cite{Wd16, k6tr}\del{)} are calling for more regulatory and legislative attention toward digital labor platforms \cite{WbPL,Lfnl,collier2017labor}. To \replace{establish effective legislation and regulation of gig platforms}{create and implement laws and regulations that effectively address key issues related to gig platforms}, policymakers and advocates require a more comprehensive understanding of the challenges and limitations that workers currently face. But due to platforms' reluctance to share related data, \add{advocates,} policymakers \add{and the public at large} currently face a data deficit when attempting to progress on initiatives for improving gig work conditions \cite{policy_probes, 2qkB}. 
% Previous research also suggests that local-level policies (e.g., state, municipal or county) wield more influence than higher-level ones (e.g., federal-level) in operationalizing and enforcing impactful regulations to protect workers and restrict platforms \cite{k6tr}, although they lack sufficient attention from the research community. However, legislation must be in place for regulatory bodies to enforce it. 
In lieu of adequate existing legislation protecting gig worker rights, our study aims to unearth the data needs of policy domain experts within the United States for 
creating and advancing legislative and regulatory policy around gig labor rights.

% creating new legislation as well as how it may inform future regulation of labor rights violations. 

% Therefore, this study aims to unearth the data needs of policy domain experts within the United States to 

% Discuss different levels of policymaking, and how local/state levels have 
% TODO: add angie's data probe for tech policy paper (chi24)
Recently, scholars identified worker-led data-sharing as a crucial step towards empowering the gig worker collective and advancing related \replace{labor laws}{policy} \cite{collectives, chiworkli, MQIo}. Calacci advocated for Digital Workerism (worker-led data-driven research and design of governance tools to shift power back to the worker) \cite{XnuE}\del{. Similarly}, Zhang et. al. suggested the use of worker data to create data probes for designing systems supporting worker advocacy \cite{WRX9} and policymaker interactions \cite{policy_probes}, while Hsieh et. al. encouraged\del{the development of} individualized policy advancements for democratizing gig work \add{across task domains} \cite{4Qr5}\add{, as well as power-aware designs for approaching sustainable data collectives \cite{collectives}}. 
Efforts by researchers and grassroots worker groups to facilitate datasharing \replace{have demonstrated a promising direction}{showed promise} for meeting the data needs of gig workers and policymakers: the Shipt Calculator \del{developed by Calacci and Pentand}helped workers track pay and combat wage theft by allowing workers to share pay data with each other in aggregate \cite{6B4U}. Fair.work allows workers to publicly rate the working conditions of platforms using surveys \cite{GgXq}, while Stein et. al. explored whether participatory design can help workers develop counterhegemonic data collectives \cite{MQIo}. \del{Especially existing bodies of work fall short in identifying concrete data needs, particularly those of policymakers, and in understanding the alignment and differences in the data needs between policymakers and gig workers.} 
% \textcolor{blue}{explicit statement of research gap here?}
However, \replace{existing bodies of work fall short in identifying}{there remain gaps amongst existing bodies of work around (1) what (shared and stakeholder-specific) initiatives and policies can worker datasharing help workers and policymakers advance (2)} concrete data needs  that policymakers and workers have for \replace{advancing policies and initiatives}{promoting such labor rights, as well as (3) practical challenges that stakeholders foresee in conducting and governing datasharing systems.}. 
% Moreover, we lack understanding around the alignment and differences in data needs between policymakers and workers.{of policymakers and workers for advancing policies and initiatives, as well as practical challenges that stakeholders foresee in conducting and governing datasharing systems.}

To understand the \add{policy priorities and} data needs of \replace{workers and policymakers}{policy domain experts and workers across domains,} as well as their \replace{preferences for}{concerns regarding potential practical challenges around} collective data-sharing systems, this study took a two-pronged approach\replace{that emulates}{, inspired by} \replace{techniques from past studies designing}{prior techniques that designed} for worker needs, preferences and wellbeing \cite{dillahunt2018designing, spektor2023designing}. First, we conducted semi-structured interviews with 11 policy domain experts from the U.S. (policymakers, policy implementers, advocacy groups, and a policy researcher) to gather policy priorities as well as feedback for designing a worker-centered data-sharing system. Subsequently, we engaged with 14 gig workers (from four task domains) in the U.S. and beyond in co-design workshops to explore and deliberate on \replace{data-sharing preferences}{whether their policy priorities aligned with those of policy domain experts, as well as preferred design choices for worker data sharing}. \del{The following research questions guided our investigation:}

% \textbf{\add{Research Questions:}}
\begin{enumerate}
\item[\textbf{RQ 1}] \replace{What are policy- and worker-related considerations, concerns, and preferences regarding data-sharing?  }{Which policy initiatives around gig work conditions are supported by both workers and policy domain experts, and which are unique to each group?}  

\item[\textbf{RQ 2}] What \add{concrete} worker-\replace{contributed}{shared} data is needed to \replace{promote policy advancements and improve individual worker conditions}{advance such initiatives}?

\item [\textbf{RQ 3}] \add{What are anticipated challenges and preferences that worker and policy domain stakeholders have regarding the practical implementation of a datasharing system?}

% \begin{enumerate}
%     \item \add{What interaction mechanisms are required of datasharing systems to sustainably engage workers across task domains?}
%     \item \add{What stakeholder groups should access, own and govern datasharing systems? }
% \end{enumerate}

% \textcolor{blue}{ON HOLD till we finish gap statement and findings reorg}
\end{enumerate}

This study contributes to the existing literature around gig worker advocacy by 
(1) expanding the community's understanding of data needs for policy initiatives and data-sharing preferences, \replace{and (2) aligning policy priorities with worker needs, so as to promote collaborations that lead toward more healthy, sustainable and inclusive futures of gig work}{(2) identifying alignments and differences of policy priorities between workers and policy domain experts, so as to further progress for future policy collaborations, and (3) mapping out practical challenges around worker datasharing as anticipated by both stakeholder groups, so as to guide and inform future designs of datasharing systems}.
Our findings show that both stakeholder groups sought 1) data to understand pay practices and (unpaid) work time and 2) more attention toward the issues of discrimination and safety. However, distinctions remain---for instance, policymakers emphasized a need \replace{for data stratified by demographic information}{to further understand and measure stressors on (care-giving) workers,} \replace{while workers}{but workers themselves} \replace{desired to learn}{prioritized learning} strategies \add{for dealing with such stressors} through the sharing and learning of qualitative experiences. 
Based on these findings \add{(and participants' stated preferences surrounding diversity, trust and ownership)}, we propose design guidelines for data-sharing systems that incorporate the perspectives of both stakeholder groups, as well as reflections on challenges that may arise around worker data integrity and policymaking.
% \textcolor{blue}{FINDINGS SUMMARY COMPLETED, DISCUSSION SUMMARY PENDING?}

\section{Related Work}\label{h.514hnj8l29rm}
% \textcolor{blue}{[overview of entire related works sections]}

\subsection{\replace{Challenges \& Inequalities Endemic to}{State of Unregulated} Gig Work}\label{h.1rbrcgukd0tl}

\replace{Over the past decade, t}{T}he widespread adoption of gig work \add{over the past decade} has been a celebrated disruption to the traditional labor market. Gig \replace{workers}{platforms} span a variety of \replace{sectors}{work domains}, including popularly recognized \del{forms of }\add{\textit{on-demand platforms} such as Uber, DoorDash or Instacart -- which leverage ``algorithmic management'' to rapidly match and dispatch low-wage workers to consumers -- as well as \textit{marketplace platforms} such as Upwork, TaskRabbit or Rover  -- which primarily sort and rank candidates for consumers who seek help with a specialized task \cite{hopwood2024demand, beyond, ticonaleft}. 
% \textcolor{blue}{however/despite differences across platforms, they share many challenges inflicted by lack of regulations}
Despite the contrasting services offered by the two types of platforms, they share many challenges inflicted by the lack of regulations \cite{policy_probes}, necessary public infrastructure \cite{nosh} and pre-existing inequalities and exploitation \cite{participation, creativity}. For instance, low-resourced workers who don't have (or know about) necessary new digital skills are disadvantaged on marketplace platforms, and end up having to perform extra unpaid learning and labor to self-brand --- an essential skill for remaining competitive \cite{beyond}. Meanwhile, workers of physical on-demand platforms are subject to performing various forms of unpaid logistical labor \cite{collier2017labor, capitalism}, in addition to increased exposures to safety and health risks \cite{rc6G, body}: rideshare and delivery drivers endure dangers of the road \cite{Krb8,jbho,8xYL}, food couriers have limited restroom access \cite{qwKX}, and cleaners and caregivers bear risks of entering strangers' homes \cite{rc6G,dCmn}. The unregulated nature of platforms subject workers to such risks without the safeguards of workers' compensation or health insurance \cite{rc6G,xMHW} and additionally suppresses their understanding of labor rights \cite{TExu}, pointing to a need for policy interventions to reshape the current landscape of gig worker protections.}  
\del{transportation-based work (e.g., ridesharing and food delivery) as well as more niche task domains like domestic caregivers \cite{BOcC}, pet caregivers \cite{mcdonald2021means}, novelists \cite{FZpg} and online tutors \cite{hdsI}. }

\replace{While gig work arguably expands work access to traditionally disadvantaged groups by offering lower entry barriers than standard employment, studies found work conditions to be rife with }{The absence of regulation has produced an extensive list of labor issues entangling gig work conditions, including rising} inequalities \add{\cite{8lEL,M7rN,eHtK,EdLY, dykp, cSUH, v6sO}}, precarity \cite{paNg,OBVk,byME}, as well as health and safety hazards \cite{PR7K}. 
\replace{Past work found i}{I}nappropriate classification 
\replace{as a leading cause for subpar working conditions and exploitation. Categorizing}{of} workers as independent contractors (i.e., non-employees) \replace{allows}{constitutes one major loophole that enables} platforms to circumvent provisions of standard benefits and protections (e.g., unemployment insurance \cite{xe4p,SPEO}, minimum wage \cite{VBof}, right to unionization and collective action \cite{VcCT,k6tr}) typically afforded to employers \cite{Yl6z}\add{, leading to poor work conditions and exploitation \cite{UwCF,Dlgz,0Ege}}. 
\replace{In the regulatory space, past work found}{Meanwhile, regulation at the} city and state levels \replace{of labor regulations to be}{remain} scarce, especially in legislative and administrative processes \del{ --- litigation is \add{relatively} more common in comparison}\cite{collier2017labor}. Moreover, state legislation (often targets of platform lobbying efforts \cite{collier2018disrupting}) frequently preempts city and local regulation, causing legal scholars to advocate for updated legal standards for the 21st century workforce and stronger labor law enforcement \cite{bernhardt2010broken}. \add{Despite widespread recognition of the lagging regulations surrounding gig work, it remains unclear which of the identified issues around working conditions lie at the intersection of (1) what are most urgent to workers and (2) what policymakers can and strive to address.}

\subsection{\replace{Data Collectives for Bargaining}{Countering Unjust Platform Practices: The Potential of Data}}\label{h.ltnlq9a3xitu}

\anedit{Amidst the lack of clear policy or regulation to protect workers or hold platforms accountable, researchers increasingly point} to the \replace{necessity}{promise} of data to rectify information \replace{gaps produced by platforms}{asymmetries} and strengthen \anedit{collective worker} campaigns. 
Khovanskaya et. al. asserted that in the absence of bargaining rights, workers need to collect their own data \anedit{as evidence of injustices (e.g., inequitable pay practices)} to advocate for labor issues such as fair wages \cite{DjK8}. 
Workers of \citet{TdXv} shared desires to engage in data investigations (e.g., collective auditing) to analyze platform incentive structures for manipulation. 
In follow-up studies, \citet{WRX9} \add{first} created data probes \replace{observing how workers used these to investigate their own rideshare data and identify experiences}{to help workers examine their rideshare data and uncover instances} of platform manipulation\add{, and subsequently explored how policy-related stakeholders would leverage them for demystifying problematic platform practices (e.g., work and wage assignment algorithms) \cite{policy_probes}}. \anedit{In a similar vein,} \citet{6B4U} engaged in worker-led auditing to reverse engineer logic behind changes in \replace{a platform's}{Shipt's} \anedit{opaque} commission determination algorithm. 
\anedit{Uniquely,} \citet{sousveillance}\del{approached the issue from a ``care ethics'' perspective to} \anedit{explored how workers may harness counter-data through \textit{sousveillance tools} to monitor those in power and increase platform transparency}. 

% This was specific to the application of rideshare gig work though, and a comprehensive understanding of policymakers' concerns around worker conditions and what data is needed to pass worker protections---especially as related to different domains of gig work---is needed.

% how to align worker initiatives with policymaker objectives: \citet{4Qr5} surfaced the types of interventions that multi-stakeholders, including regulators, envisioned for gig worker well-being; and
%In addition to challenges of occupational hazards and inequality, the individual nature of gigs leads workers to feeling atomized and isolated \cite{Hv7i,k6tr,ERsM}, making it difficult for them to collectively make demands or advance initiatives. Such isolation arise from intentional design of platform designers, who withhold information around algorithms from workers to produce power asymmetries and induce work coercion \cite{GBkH}. 

\anedit{While data holds potential for supporting worker protections, current tools for data collection and analysis advance worker-centered policies and initiatives in limited ways. For example, while third-party developers have created apps to help gig workers track their work data, these center around individual tax reporting purposes (e.g., Gridwise, Stride, and Everlance all help workers log metrics such as mileage, jobs worked, and expenses incurred). Meanwhile, researcher-created tools are often narrowly scoped (e.g., to one platform \cite{6B4U}) or aim to help workers surface (collective) concerns but lack strategic alignment and features to directly influence policy change \cite{WRX9, weclock, savage2024unveiling}. Thus, we currently lack an understanding for how tools can (1) support scalable harnessing and aggregation of worker data for informing related worker initiatives and (2) align such collected data with policy and regulation to enable wider-reaching change.}

\subsection{\anedit{Towards Data Collectives that Respect Needs \& Workflows of Laborers Across Task Domains}}
%%put on hold until after findings 11-20-2024
%%first talk about considerations from other domains about sustainable and large-scale data collectives; also include jakes exploration
%%then focus on how non-rideshare domains are underexplored and present different challenges

%%considerations for sustainable and large-scale data collectives 
\anedit{To approach helping workers harness data for initiatives while ensuring alignment with policy and regulation capabilities, we draw inspiration from recent research around collective data contributions and data donations. While not specific to gig work, \citet{GUVs} and \citet{a9kx} suggested the potential for technology users to leverage their data and resist company practices (e.g., privacy infringements, biases in AI systems), by facilitating ``conscious data contributions'' wherein individuals donate data towards a specific campaign.}
\anedit{Relatedly, the concept of ``data donations''---user contribution of their own data for \textit{academic} research---is also being explored, in domains like} social media \cite{razi2022instagram, zannettou2023leveraging, meyer2023enhancing} and healthcare \cite{strotbaum2019your, bietz2019data}. Research here primarily focuses on understanding user motivations for donating \cite{keusch2023you}, potential selection bias of donations \cite{kmetty2023determinants}, and infrastructures for secure and trustworthy collection \cite{boeschoten2023port, zannettou2023leveraging, carriere2023best}. 
% \anedit{To surface considerations for how to harness data for worker initiatives and ensure alignment with policy and regulation efforts, we turn to complementary research about how to help individuals contribute data towards a collective goal. Here, some researchers suggest the potential for technology users to leverage data to} resist problematic company practices (e.g., privacy infringements, biases in AI systems), such as \anedit{\citet{GUVs} suggesting organizers initiating "conscious data contributions" where people donate their data towards a specific campaign for change} \cite{a9kx,GUVs}. 

\anedit{Importantly, the gig work context presents additional elements to factor in when considering data collection, such as a diversity of task domains (each involving its own unique set of data types), a broad range of labor issues and initiatives where aggregated data can be applied, and several involved stakeholder groups --- both platforms and consumers hold power over and collect data from workers, risking worker (and possibly consumer) privacy and agency.} \add{A multi-platform social media analysis from \citet{sannon2022privacy} found gig workers to experience intrusive data collection and surveillance not only from platforms but also customers. 
% In particular, Vashistha et. al. revealed how contextual factors can shift how resource-constrained users engage with technology, and thus also change their privacy needs and preferences \cite{vashistha2018examining}. 
Around policy development, \citet{kahn2024expanding} showed how privacy concerns of impacted communities deviated what's expected by privacy and development experts, suggesting a need to (re-)align preferences of higher-power groups with those of experiential experts.} 
\add{To more comprehensively understand practical implications of worker data contributions (e.g., privacy concerns, power dynamics with consumers/platforms, and (un-)intended impacts of policy developments), it is imperative to involve workers 
when making design decisions around worker data contributions, especially when such data are meant to eventually impact policy.} 

As a first step in this direction \citet{MQIo} held co-design sessions and surfaced rideshare drivers' \anedit{preferences for contributing, collecting, and using data---in other words, }collective data infrastructures---including ``collective wikis'' and ``new app'', {involving a simple, separate application with mechanisms }to support data collection, sharing, and governance. However, \replace{much of this work focused solely on rideshare or delivery drivers}{by only engaging with impacted drivers}, \replace{thereby excluding}{this approach risks sidelining the preferences of} other worker groups or stakeholders \add{who are, or may become,} involved in worker advocacy. \del{While }Hsieh et. al. {engaged with workers,} advocates, regulators, and platform employees {to surface} priorities \replace{that each group held}{of each group} around \del{gig }worker rights, \replace{their workshops did not focus specifically on designing}{but these workshops covered a broad space of policy, service and technology solutions, instead of focusing on} collective datasharing infrastructures \cite{xAXX}. 
\add{Furthermore, the expanding diversity of gig work domains call for closer examinations of how regulation can be improved across sectors \cite{4Qr5}, especially since risks and responsibilities vary widely across platforms \cite{sannon2022privacy}, and the lack of governance between occupations can differentially impact how workers across sectors experience such risks \cite{beyond, mastracci2016breaking, basu2015health}.}
{This study extends these works to \replace{explore workers' preferred data institutions  (particularly the ``new app'') for facilitating knowledge exchange, as well as to examine how data-sharing systems can support policymakers' initiatives for improving gig work conditions across task domains.}{codesign for worker data exchange and knowledge sharing in a way that meets policy priorities and data needs of workers (across domains) and policy related stakeholders.}
}

% \textcolor{blue}{in sum, lack of understanding clarity -- summarize three gaps identified in three sections, thus in the study we do x to bridge the y gap}
\add{In sum, the existing bodies of work has yet to identify (1) policy and regulatory advancements around gig work conditions prioritized by workers and policy experts, (2) how worker datasharing can scalably support such advancements and (3) ways of accounting for practical implications of 
datasharing by workers across a diversity of domains/platforms. This study aims to bridge these gaps by engaging with both policy experts and workers across four gig domains/platforms to identify their shared (and misaligned) priorities, data needs for meeting such priorities, as well as considerations and preferences for practical impacts of worker datasharing across multiple platforms.}

% \section{Research Gap}

\section{Methods}\label{h.5sn3xgklbkyx}
To inform the design of a data-exchange platform, we conducted semi-structured interviews with policy domain experts to gather which aggregate worker statistics are useful to advancing policy for gig work. Then, we led co-design workshops with gig workers to understand their motivations and concerns around collective data sharing, {following the precedence of works that examined hospitality work and underserved job seekers \cite{dillahunt2018designing, spektor2023designing}.} {With policy domain experts, we chose interviews to focus on open-ended discussion and discovery about their policy efforts. 
For gig workers, we deliberately held co-design workshops to allow space for participants of various backgrounds and experiences to share preferences and ideas together, so as to allow workers to collectively deliberate on design decisions for a data-sharing system that meet collective goals of a diverse worker population.}

\subsection{Recruitment and Participants}\label{h.kvn6j5yejjmz}
\subsubsection{Stage 1: Interviewing Policy Domain Experts}\label{h.pu0gzew4e1gn}
{Through contacts from the research institute Metro21, w}e recruited 11 over-18, US-based policy domain experts to semi-structured interviews (Table \ref{tab:pde}). For this study, we consider anyone who makes or influences policy as a policy domain expert. {To gather both local and national-level insights, we intentionally recruited participants from city-, county-, and federal-level offices.}

\subsubsection{Stage 2: Co-designing with Gig Workers}\label{h.vchog3sn8mse}
For gig worker co-design workshops, we recruited 14 US-based, over-18 active gig workers through a combination of Reddit posts, word-of-mouth, and  previous study participants (Table \ref{tab:worker}). Using a pre-screener survey, we recruited workers based on types of gig work and demographics: rideshare and delivery drivers to represent more popular sectors of gig work while petsitters and freelancers account for perspectives of those performing more niche tasks. Demographically, we intentionally oversampled from underrepresented populations to explore the impact of intersectional identities, and no information was collected regarding participants' prior experiences of sharing data with other researchers or organizations.

% \FloatBarrier
\begin{table*}[h]
  \centering
  \includegraphics[width=.95\textwidth]{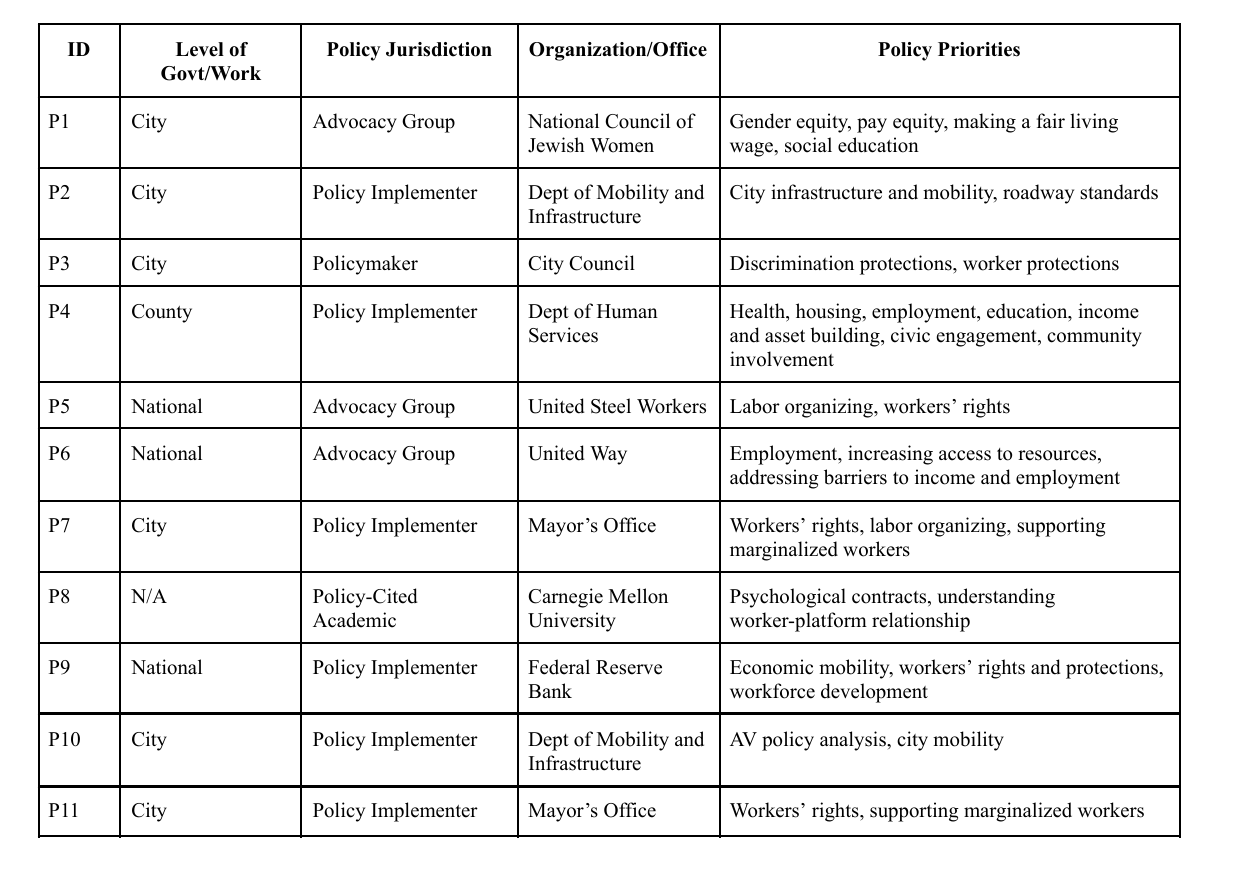}
   \caption{Overview of Policy Domain Expert Participants}
  \label{tab:pde}
\end{table*}
% \FloatBarrier

\subsection{Protocol and Study Design}\label{h.76kn6ynhoa9i}
\subsubsection{Stage 1: Interview Protocol with Policy Domain Experts}\label{interviews}
As policy begins to govern data and AI systems, we sought to understand how a worker-centered data-sharing platform would impact public infrastructure while bridging the research-policy gap. Each of the 11 semi-structured interviews involved one to two policy domain experts, and was organized into general themes of 1.) understanding interviewees' goals and policy processes, 2.) data to support interviewees' decision-making, and 3.) platform and data governance. To help participants generate ideas and make decisions on relevant worker data, we prepared and presented a list of potential data types, divided by occupation (e.g. rideshare, petsitting). {For example, rideshare driver participants saw data types including time/date/location of trips, driver wages, and total paid by passenger while freelancers saw data types like Job Success Scores or customer review ratings.} {We asked interviewees} to 1.) decide whether such data would be helpful and 2.) generate additional pieces of useful worker data and 3.) share rationales of how they would leverage the data. Finally, we ask for participants' opinions about the best way to manage the platform from a policy perspective ``without sacrificing data producers' [gig workers'] control'' over their data contributions \cite{GUVs}. 

\subsubsection{Stage 2: Study Design with Gig Workers}\label{h.5r3oji4mg7tb}
With workers, we held a total of four co-design workshops (with 3-4 participants per session), where each focused on a specific type of gig work: freelancing [F], food delivery [D], ridesharing [R] and petsitting [W]. 
Sessions lasted 90-120 minutes, and participants were compensated at \$60/hour{. We} verified active gig working status through screenshots or live showings of {profiles}. 
Each workshop consisted of four sections focusing on 1.) incentives for data sharing, 2.) types of {useful data}, 3.) sharing preferences, and 4.) platform and data sharing concerns. 
To understand worker experiences and data needs, we formed new questions with participants {after introducing sample questions and issues generated based on previous findings \cite{xAXX,4Qr5,TdXv}}. 
Then, we kept workers' answers to experience- and need-related questions in mind as we collectively probed {their desired stakeholders and data types to contribute}. 
At the end (to prevent priming workers' responses to earlier topics about data), we introduced initiatives \emph{informed by the previous interviews} (\ref{interviews}) to learn more about {participants'} thoughts, opinions, and experiences as workers across various intersections. 
% \anedit{Initiatives were presented at the end to prevent priming workers' responses to earlier topics about data.} 

For each workshop activity, participants created sticky notes on a Miro board and ranked these stickies alongside prepopulated notes. Rankings were determined on a rotation of participant emoji reactions, clusters on a linear scale, and groupings via quadrants. All prepopulated data was specific to the type of gig work each workshop focused on. We share our protocol  and study materials in supplementary materials. During the workshops, we collaborated with workers to identify data types they felt comfortable sharing, their preferences for methods of sharing (via file uploads, automated processes, etc.) as well as concerns and reservations against data-sharing.

% \FloatBarrier
\begin{table*}[]
  \centering
  \includegraphics[width=.89\textwidth]{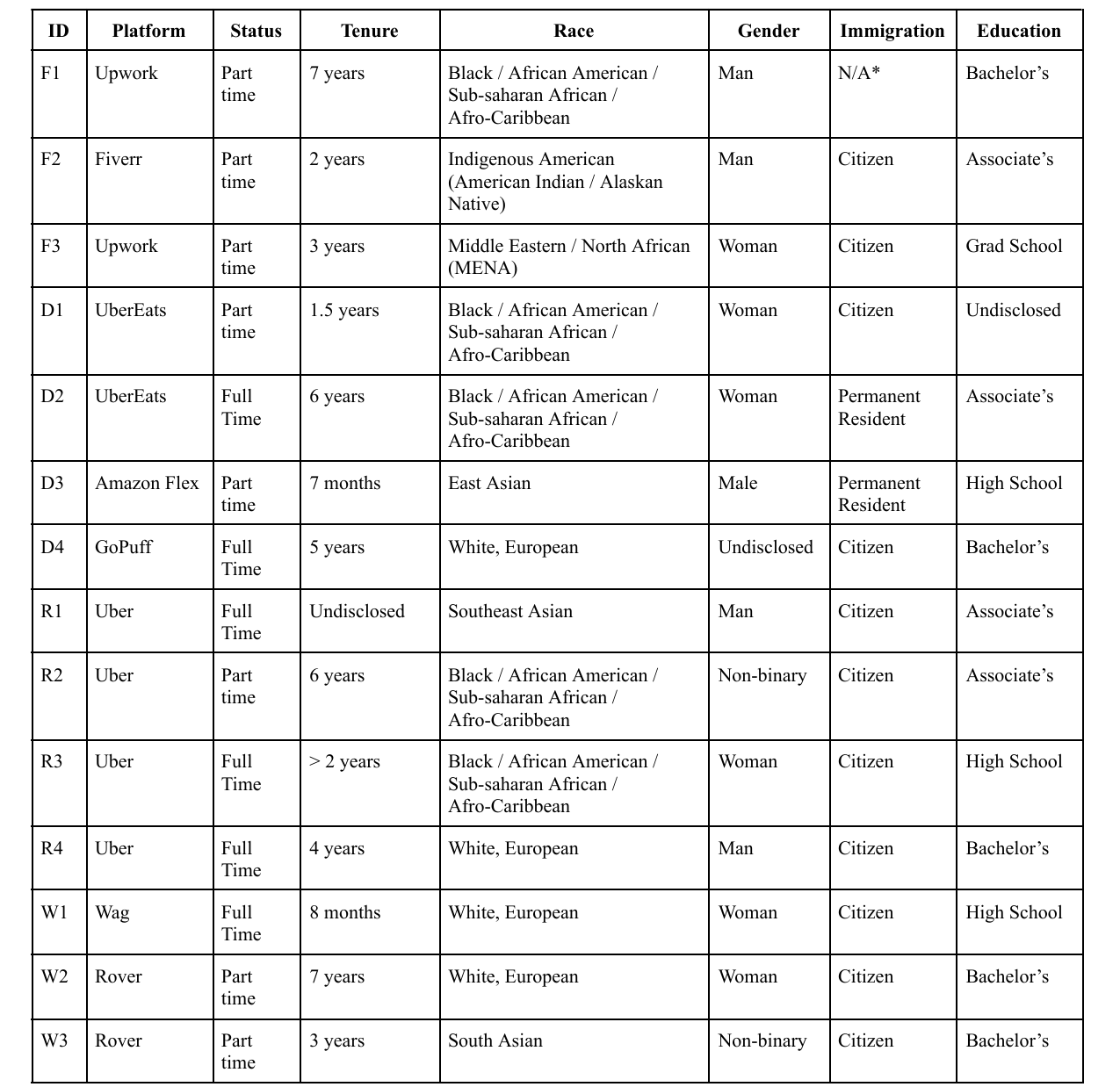}
   \caption{Gig worker participant demographics}
  \label{tab:worker}
\end{table*}
% \FloatBarrier

\subsection{Positionality}\label{h.m6dj693qlhxl}
We recognize that our work is directly informed by our experiences and positions as designers and researchers, which provides us privileges our participants do not possess. In particular, our team members work and receive training in the United States in the fields of Human-Computer Interaction, Software Engineering, Computer Science, and Public Policy. We hold prior research experience in human-centered computing, with four authors having direct prior experience engaging with workers and policy domain experts in participatory design contexts. However, our prior experience \emph{does not} remedy our advantage as researchers within the power imbalance of the research-participant relationship and no one in our team has participated in either the gig economy or policy-advocacy space. 

To ensure a worker-centric procedure, we consulted closely with gig workers as domain experts. As recommended by prior literature, we openly disclosed our intentions to aid workers and positions as researchers working independently from gig platforms with participants before all sessions. Furthermore, we continuously reflect on three main questions inspired by \cite{cKW0}: (1) What biases are we bringing into this space? (2) Are we falling into a savior complex? Or are we truly supporting and serving the gig work community? (3) Ultimately, in what ways does this work change realities for workers? We also took special consideration to how our participants could benefit from our work outside compensation as time is precious, especially in the case of gig workers. 

\subsection{Analysis}\label{h.yglm2551dqq7} All interviews and workshops were recorded on Zoom and later transcribed using \emph{Rev.com}. We used a qualitative thematic analysis approach \cite{WMCa} to analyze both the interviews and workshops. To begin analysis, we adopted an open coding approach on the interview and workshop transcripts where one researcher independently generated codes, applied at sentence or paragraph levels \cite{C5bR,yWbe,WMCa,CQOi,T92p}. At least one other team member then cross-checked the initial codes for each interview and workshop transcript, where no less than one of the coders was present in the corresponding session. During this process, both coders remained receptive to uncover as many new codes as possible, while keeping in mind our research questions around worker data sharing preferences and policy advancements for improving work conditions. Coders met on a weekly basis to 1) develop a system of assigning IDs to participants while preserving anonymity and 2) discuss and resolve any disagreements about the initial codes. Next, we iteratively combined the resulting 1593 (1022 interviews \& 571 workshops) unique codes into thematic categories, wrote descriptive memos, and built an affinity diagram from the bottom-up to draw connections between categories \cite{uTeY,sQQh}. This analysis generated 118 first-level themes, 17 second-level themes{ and} four third-level themes, which we report on {below}.

\section{Findings}

\add{Policy domain experts expressed interests in various prioritized initiatives --- e.g., expanding equitable access, approaching fairer pay practices, and reducing work-induced stress --- as well as excitement about worker-based data sharing, describing how worker statistics can help advance related policies, especially if they are aggregated to better inform funding resources and program development for workers (P1, P6, P7, P11) and is also less prone to manipulation by ill-intentioned actors (P5). Workers similarly advocated for safety, equity and fair pay practices, but they additionally yearned to learn experiences and strategies from other workers -- this inclination toward qualitative datasharing matches the desire of policy stakeholder participants, who
wanted to leverage \textbf{qualitative accounts} to help them understand job quality and workers' stories (P1, P7, P8, P9). }

\begin{figure}[h!]
  \centering
  \includegraphics[width=\textwidth]{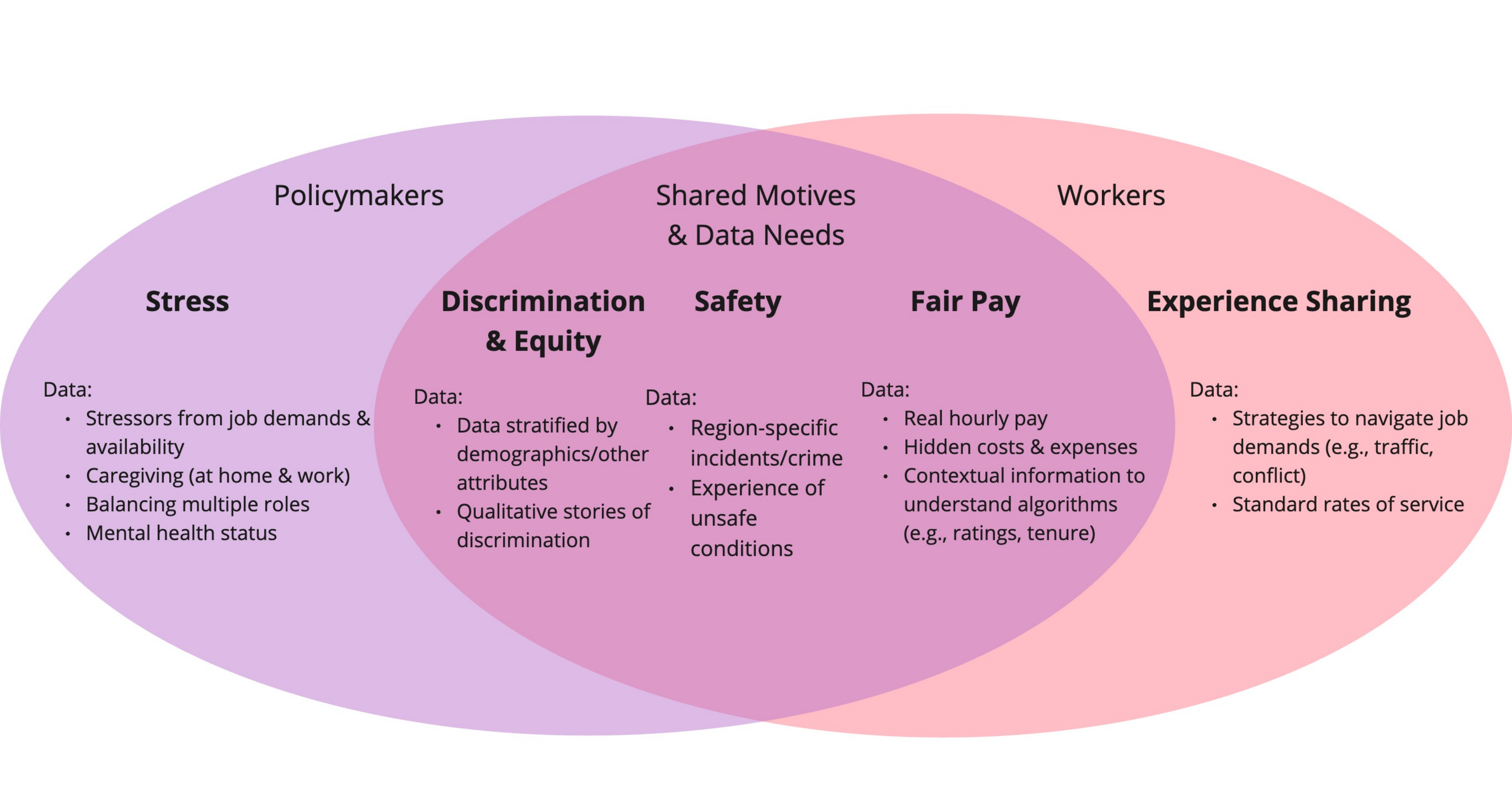}
   \caption{Summary of Main Findings: Figure shows initiatives that policy domain experts and workers desired to support with data collected through a data-sharing system. Center of diagram demonstrates three shared initiatives between the stakeholder groups.}

\end{figure}
\FloatBarrier

\add{We organize these findings by first presenting the three initiatives of interest shared by both stakeholder groups (\ref{shared}), followed by interests that primarily pertained to policy domain experts (\ref{stress}) or workers (\ref{h.pplzn44x8y2i}), and lastly the practical preferences and anticipated challenges that participants expressed around future worker datasharing collectives (\ref{h.tycl8ab0p8ie}). In each section, we first discuss results from interviews with policy domain experts, followed by insights gathered from workshops with workers --- to remain consistent with the chronological order of how the the study was conducted.}

\subsection{Initiatives \& Data Needs Shared Among Stakeholders}\label{shared}

\subsubsection{Equity \add{\& Discrimination: (Dis-)aggregated Understanding of Traditional (Demographic) \& Domain-Specific Factors}}\label{pde_disc}
\paragraph{Policymaker Rationales for Prioritizing Equity}\replace{Government officials that we spoke with }{Policy domain experts }most commonly prioritized equity among other initiatives, both locally and nationally. At the city level, \add{P3 (a local council person)} mentioned their work {expanding} on existing national anti-discrimination policies: ``already in most cities across the country, it's illegal to discriminate on the basis of sex, race, religion \del{, many others.} so these \del{[new protected classes I work for]} just add on to the list \add{[of protected attributes I would advocate for]}''. At the national level, P2---whose office works to influence federal policies---{shared} how ``the federal government right now also has a very strong equity focus.'' Equity was brought up with respect to \textbf{several protected groups}\replace{ (e.g.,}{:} female, disabled, \add{and undocumented} workers\del{)}. 
\add{Both P8 and P7 worried about the potential for gig platforms to exploit already vulnerable populations}, especially since ``a lot of people who are gig workers are in marginal positions'' (P8), \add{while P9 sympathized with gig workers whose lack of English fluency became a ``negotiating tool by employers to underpay or sometimes not even pay them.''}
\add{To protect female workers,} P1 specifically described how their organization examines ``gender equity--- understanding specifically women who are doing part-time work,  \dots the demands \dots where supports are lacking for women'', whereas
members of the mayor's office recounts how ``the city has historically really looked from a procurement standpoint around minority and women-owned businesses'' (P7). 
Meanwhile, P6 works to provide support to ``workers with disabilities  \dots  that either are working or get any Medicaid or assistance programs''. 
\del{Members of the local mayor's office recounts how ``the city has historically really looked from a procurement standpoint around minority and women-owned businesses'' (P7). Policymakers also worried about the potential for \textbf{gig platforms to exploit already vulnerable populations}. }
\del{For example, P7 argued that while the low barriers of gig} \del{ platforms gives work access to traditionally marginalized groups such as undocumented and uncertificated citizens, platforms are still rife with SES disparities since they often exploit workers with short ``planning horizons'' by leveraging their need for temporal flexibility. }
\del{P9 similarly sympathized with gig workers whose lack of English fluency became a ``negotiating tool by employers to underpay or sometimes not even pay them.''}

\paragraph{Worker Concerns around Discrimination} \del{Discrimination was reported among a}Worker experiences reflected concerns around traditional forms of discrimination, as well as factors unique to the with each \del{type of }work \add{domain}: workers from matching-based marketplace platforms (freelancing \& petsitting) worried about common factors for biases such as race, gender and sexual identities while those providing rideshare services described the potential \add{domain-specific} impacts of car models. 
\del{Freelancers and petsitters worried about more traditional factors: \textit{\textbf{race}}, \textit{\textbf{gender}}, \textit{\textbf{sexual identities}}, and \textit{\textbf{age}}. Food couriers described how \textit{\textbf{spoken languages}} and \textit{\textbf{immigration status}} may help or hurt earning prospects while rideshare drivers interestingly relayed inequitable treatment based on \textit{\textbf{car model}}.} Freelancer\del{s observed the most egregious case, with} F1 report\replace{ing}{ed} ``a lot of \replace{''discrimination on platform and that some clients ``}{ [cases with clients who] }actively discriminate \dots [by putting up] job posts \replace{where the clients }{that} say ‘no Indians, I don't want Pakistanis, I don't want people from Southeast Asia'  \dots   and some others could do it in terms of the choice of freelancers that they hire''. 
Food couriers \add{like D1} worried about impacts of \replace{immigration status and the number of languages spoken. On the one hand, D1 noted that}{sociocultural factors:} ``speaking multiple languages helps a lot because  \dots  down the road you'll interact with several customers and it matters a lot'', \replace{yet}{while} D2  \replace{speculates that immigrants may be competitively disadvantaged}{points out the disadvantages of immigrants face:} \del{because} ``if \dots  based on your immigration status you do not have that driving license, you have to use another means (e.g. a bicycle) \add{[but]} then that means you wouldn't pick deliveries that are long distance.'' Petsitters like W5\del{ also worry about} \add{ expressed similar concerns about} discrimination \add{based on demographic factors} \replace{ and}{since }``I have a visibly ethnic name, [so] I sometimes wonder whether experiences with these platforms  \dots  are affected by  \dots  being perceived as foreign or frankly not being white  \dots [since] the majority of clients are white, especially when you get into more affluent neighborhoods.'' W3 \replace{similarly wanted}{also wondered about} beginner petsitters who ``have an ethnic name  \dots  visibly non-white or visibly queer or out as trans on your profile [to know]: Is it statistically indicated that your earnings are going to be less, or your safety is going to be more compromised?'' while W2 \replace{wondered}{questioned} ``if you are part of the LGBTQ community, how does that affect your decision making and \dots how people hire you?'' %Another factor biasing the petwalking population is age, which ``typically skew towards a somewhat younger sitter. A lot of these people  \dots might not have a laptop  \dots doing this entirely via mobile.'' 
Rideshare drivers uniquely mentioned \replace{that car conditions can }{the differential }impact\add{s of car model and year} \replace{discrimination by both the}{on both} platform and clients. Speaking to platform discrimination, R1 claimed that ``if you have a newer year or a newer model, of course they [the platform] prioritize [\replace{job allocation}{rides} to newer cars] over an older car''. \del{Speaking to client discrimination, }R4 \replace{shared that}{offered an example where} ``my buddy just started renting through Tesla  \dots  he's seen a significant increase \dots \del{making about a little bit more,} even with the \$600 a week rental charge''. %Such observations made R4 ponder: ``the first person who gets the ride, they have the newer car  \dots  [then] if the second person and third, fourth person get rides, is the rate from the first person higher than the other three on average?''

\paragraph{\add{Ways of Understanding Equity \& Discrimination with Worker Data}} \label{data_disc} \replace{Participants'}{The} equity concerns \add{shared by policy domain experts} motivated them to use data for investigating whether discrimination is occurring to gig workers \replace{. They explained the importance of viewing }{at the aggregate level, as well as how such discrimination breaks down across }\del{data stratified by }demographics to understand whether vulnerable groups --\del{particularly}ethnic and gender minorities\replace{ and }{, as well as }those with disabilities (P3, P6){-- are experiencing the impact of discrimination disproportionately. P2 and P5\replace{explained that analysis around}{, for instance, suggested analyzing} gaps in compensation and work assignment across demographics \del{can be  used }to measure discrimination\del{ of platforms} \replace{. And}{ while} P2, P3, and P5 mused whether data such as \del{passenger's rating of a }worker \add{ratings} and worker's \add{acceptance/cancellation} rates\del{of acceptance or cancellation} could reveal racial prejudice of customers against drivers: ``I've seen research showing that when people get information, sometimes they'll cancel things based on driver's background \del{\dots whether it's due to prejudice on the end of the driver or on the client}'' (P2). P5 \replace{were also curious about}{contemplated} the mechanics of algorithms: ``Are there ways that the work is either assigned or accepted or offered that are discriminatory?'' Relatedly, P6 and P7 wanted to assess whether negative customer reviews may coerce workers into accepting unsafe jobs or conditions, which may require narrative accounts from workers. 
\del{Finally, to gain insight on the equity impact of gig work, participants discussed how data could help them understand the impact of gig work on equity. For example, P7 and P11 discussed using a worker's travel time as a proxy for worker's SES to see whether there is ``a disconnect between the background of the individual and the location of where they're providing the service'' (P8). }

% \subsubsection{Discrimination}\label{worker_disc}
%angiee working on this 6-21-2024 (keeping most of the same bc it's already concise)

% \paragraph{How Data Can Support \& Provide Insights on Discrimination:}

While \del{policymakers are focused on understanding potential disparities at an aggregate level, our }gig worker participants \replace{were 
interested in learning about the }{showed similar interest in learning the aggregate }demographics of their fellow workers\replace{ and their individual}{, they also yearned to hear about their peers' personal and anecdotal} experiences with discrimination. For example, D1 \replace{wanted to understand}{wondered} how \del{their}\textbf{spoken language capabilities} of drivers impacts jobs. 
D2 believed presenting policymakers with \add{disaggregated} statistics on worker safety and pay disparities \replace{that are disaggregated }{(}by demographics such as {race and gender}\add{)} could make evident patterns of discrimination\replace{ and potentially}{, which may} lead to policy action. 
\add{R4 imagined an experiment where four co-located driver friends all made themselves available on the app at the same time: ``the first person who gets the ride, [if] they have the newer car  \dots  [then] is the rate [of receiving rides] from the first person higher than the other three on average?'' With metrics on ride offers received by workers of similar/different backgrounds (or similar rides by drivers of different backgrounds) can help identify, elucidate and alleviate the discriminatory impacts of (on-demand) gig work.}

\del{R4 also gave an interesting example of how scenario for this might play out to investigate if platforms are discriminating based on a driver’s car type: ``the first person who gets the ride, they have the newer car  \dots  [then] if the second person and third, fourth person get rides, is the rate from the first person higher than the other three on average?'' Data about offers drivers receive for the same/similar trips could be used to analyze for discrimination on attributes such as worker's car type and demographic attributes. From metrics like this, gig workers hope one could identify and alleviate discriminatory behaviors in gig work. }

\subsubsection{Fair Pay\add{: How Data Transparency Can Take Workers from Subminimal $\rightarrow$ Livable $\rightarrow$ Fair \& Profitable Wages}} \label{pde_pay}
\paragraph{Why Policymakers Focused on Fair Pay}
\replace{Participants}{Policy domain experts} also stressed the importance of fair pay, a criteria requiring (1) workers' earnings to meet a livable minimum wage and (2) be appropriately compensated for their services and expenses incurred along the way. P4 related using income to assess workers' self-sufficiency: ``we do have a measurement of living wage work, so workers\del{ or participants} [can determine]  \dots  if they're earning income that's under a living wage.'' P1 also explained how their organization's ``primary focus on the policy side has really been on \textbf{fair pay} -- (i.e.) creating an environment where folks who are working in a gig economy can make a living based on the work that they do, and make a fair wage.'' 

The \textbf{practice of tipping} is \replace{a}{one} subtle way of \replace{impacting}{undermining} pay equity \add{that policy domain experts worried about}, especially for traditionally disadvantaged groups. P2 expressed concern that ``some of the argument for having minimum tipped wages in places is because of racial and other variables that influence tips beyond the service that was provided.'' \add{In alignment with recent literature \cite{5qBZ},} tipping can also condition women into tolerating \del{mistreatment like }harassment \replace{. P7 posed a potential dilemma where if }{-- P7 posed a dilemma where if a client} ``\del{somebody}wants to pat you on the butt and you're a woman – are you going to say ‘knock it off,' or are you going to say ‘Eh, I can probably skate there. I think my tip is going to be bigger'.'' Ambiguities associated with tipping on platforms can aggravate biases – P7 expressed frustration that ``I don't know what [are] rewards and punishments from the platform in tipped work.'' 

Invisible labor is a well-documented phenomenon among gig work \cite{l2xo,QNoU,EGjj,9Fa7}, and thus \textbf{wage theft} (i.e., the practice of underpaying work or not paying workers for certain parts of completed labor) is a covert but direct \replace{way to exploit workers}{method of labor exploitation}. 
Although workers are aware of the risks inherent to platform work, they remain helpless during times of financial precarity, especially when they bear many undocumented and indirect burdens such as ``the expense and the cost to them as an individual to maintain a car'' (P9). 
%P9 related being ``very curious  \dots  about wage theft – so instances of people being either underpaid or not paid at all for duties performed''.
%, after hearing instances where ``I'm providing someone childcare or I'm doing something in someone's home, and that I'm being told that I will not be compensated for my work''. 
The integrated and often undocumented nature of caregiving work can subject workers to \textbf{invisible and unpaid labor}. P9 recalled ``specifically of a person who talked about a care job where they were not compensated,'' and emphasized the difficulty of navigating such situations – ``what happens in that instance? How do you have any ability to negotiate? Do you go to a third-party resource? \dots There is so much \dots domestic and care work that \add{[doesn't get paid, because it]} is essentially [a] gig that isn't necessarily 'I'm in this person's house every day' but is ‘every now and then I pick this up or I pick that up'.'' 

\paragraph{Worker Perspectives around (Un)Fair Wages}\replace{Workers raised concerns around how platform practices of algorithmically determining wages (i.e. algorithmic pricing \cite{kloostra2022algorithmic, pignot2023pulling}, which (re-)calculates pay based on data obtained through intensive worker surveillance \cite{dubal2023algorithmic, teachout2023algorithmic}) often result in aggressively low and sub-minimal wages.
}{Workers explained concerns about sub-minimal wages due to platform practices of aggressively low and often algorithmically-determined prices. 
Recent literature has characterized this profit-maximizing strategy \cite{dubal2023algorithmic} as \textit{algorithmic pricing} wherein {platforms use machine learning algorithms to (re-)calculate prices based on a number of parameters \cite{kloostra2022algorithmic, pignot2023pulling}, including data about workers obtained through intensive surveillance \cite{dubal2023algorithmic, teachout2023algorithmic}}. 
i}
\del{This strategy allows platforms to maximize their profit at the expense of worker’s wages and well-being \cite{dubal2023algorithmic}. }Petcare and driver participants described two \replace{types of algorithmic pricing tactics imposed by platforms}{instantiations}: upfront\del{ pricing} and dynamic pricing. %: algorithmic pricing tactics. % and inconsistent, often deceptive payment practices.

Driver participants explained \replace{wage calculations were once based on time and distance expended upon task completion, but then}{how wage calculations were once calculated with time and distance expended upon task completion, before} platforms introduced \textbf{upfront pricing}, which assigns rides with predetermined compensation, in name of improving transparency. However, drivers \replace{believe}{hypothesized that} this enables platforms to vary their \del{own }commission rates and assign sub-minimal wages, \replace{observing }{citing observations of }drops in overall compensation for same or more completed work. R4 explained, ``Now they give you an `upfront price', which is typically extremely, much, much lower than what the customer paid  \dots  around 60\%, is what Uber takes  \dots and they [dis]guised it through the lens of, `oh, you're gonna get to see what you get per ride ahead of time.' '' Not only are upfront prices sub-minimal, \add{workers are also rarely compensated for their extra efforts }if the task takes more time or travel than original \replace{ly estimated by the platform}{estimations }\del{, workers are rarely compensated for the additional effort }(R1). 

\textbf{Dynamic pricing} is another feature platforms use to arbitrarily change prices based on demand and supply --- leading to worker concerns around wage fairness and exploitation. 
D2 protested ``sometimes it might not be fair  \dots maybe there is lots of work, but the pay is a bit low because they might have used the demand strategy'', which makes them wonder ``how might the platform be exploiting workers or customers through dynamic pricing?'' R4 worried platforms could even be enacting wage ceilings on workers, believing ``the dynamic pricing and their fees on average will not let you make more than \$36 an hour'', although currently they ``don't have a way to validate that data point''. Dynamically-determined prices can also \replace{be misleading:}{cause misleading estimates:} Petsitter W1 explained \replace{that platform}{how} earnings projections \del{on tasks }can differ from actual wages received: ``[the algorithm will say] it costs \$16 for a 20 minute walk, but then it's really \$10 [or] it's \$36 for an hour walk, but then we get \$21.'' 

Finally, workers vehemently call for policymakers to provide \textbf{accessible education about the financial risks} and responsibilities of gig professions \replace{and provides clear \textbf{guidelines on worker boundaries and rights}. Currently, there is a significant}{to mitigate the current} lack of financial \replace{knowledge}{understanding} amongst gig workers around income structure, taxation, and metrics they should track and monitor\del{for financial well-being}. This knowledge gap enables platform-side exploitation, as ``a lot of gig work in general prey on people not being financially educated or not being able to forecast what is my actual earning going to be from this? How am I going to set aside money for taxes on all of this?'' (W3).

\paragraph{\replace{Data needs for}{Ways that data can reveal insights around} fair pay: }\label{h.k49wp92meqn4}
\replace{Participants}{Policy domain experts} wanted data to understand {whether workers are making a \textbf{livable wage} through gig\replace{ work}{s}} and to what extent the workers are \textbf{compensated appropriately}\del{ based on the work they do}, often referencing a desire to view how many hours (P1, P6) and how many jobs (e.g., multiple platforms or jobs) gig workers have to take on to make a sustainable wage (P1, P4, P6, P7, P8, P9) or ``get to a certain threshold of desired income'' (P4). In particular, participants wanted data to explore \del{three components of fair pay:} 1) the hidden costs and expenses workers assume themselves and how it impacts earnings \replace{,}{and} 2) the amount of paid time compared to actual time spent working\del{, and 3) how the flexibility that workers experience in on-demand work is offset by the precarity of work availability and wages}. Around \textbf{expenses}, \replace{participants}{policymakers} wanted to know what \replace{costs workers incur}{types of costs are incurred} and how they cut into \del{their }net earnings (P1, P2, P6, P9). P1 suggested using this information to better educate people about whether gig work can earn a livable wage, and P9 wanted to assess these metrics of real work profits against platform claims about potential earnings: ``[Platforms]'ll be like, you could make this much a week, but we're not accounting for the costs.'' In terms of \textbf{working time}, participants were highly concerned about \replace{the extent to which}{how much} workers spend on unpaid labor: P5 suggested \replace{collecting data on the amount of extra work for}{measuring the extra labor performed by} freelancers that are not stimulated in an original contract, as well as the scoping work required to procure a contract. For rideshare drivers, P9 mentioned \replace{gathering data on}{documenting} how much time workers spend commuting to or waiting on a passenger. Regarding the unpaid time of delivery drivers, P2 mused, ``If they're theoretically making a lot on a delivery but they have to idle and drive around for 30 minutes in between, how good of a gig is that?'' 

Workers \add{primarily} suggested \replace{several uses of}{using} data for \replace{overcoming two challenges}{achieving two goals} related to wages: 1) deciphering algorithmic pricing tactics of gig work platforms, 2) calculating metrics to understand\del{ their} earnings and devise strategies for improvement \add{and 3) spread awareness and advance safeguards around low rates (and the consequent long hours)}. First, workers envisioned combining \textbf{contextually specific information} alongside historical data \add{to understand algorithmic pricing}. For example, W2 suggested gathering data about how many other workers were solicited for the same task because the ``[platform] suggests that you reach out to more than one sitter after you message the first sitter'' \replace{. This could be  used to}{-- such stats can help workers} reverse engineer whether/how platforms use demand and supply data points to set upfront or dynamic prices. Historical work data can also support D3’s suggestion to analyze how dynamic prices vary during times of high demand. Second, \add{workers described leveraging data insights for furthering their understanding and improvements around earnings,} given the high likelihood of platforms continuing with algorithmic pricing\del{, workers described other insights and data to support them in understanding and improving their earnings}. F1 and W2 envisioned a \add{transparent} system for ``tracking your income, the fees that you're paying and taxes'' so workers \del{have pay transparency and }can ensure they are not ``running the entire operation at a loss''. Related income variables would include a diverse array of \textbf{non-financial information} (e.g., number of jobs completed, completion rate, reviews, cancellation rate, acceptance rate, etc) alongside standard pay and tipping rates.  

Workers also suggested specific metrics from a data-sharing system to help them assess \del{their }profitability, such as F3’s wish to view ``hourly rate of service'' and R4’s desire to see average pay broken out by attributes including ``ride, ride type, tenure, etc''. Multiple workers also emphasized the importance of \textbf{ratings and reviews} to secure future work, thereby ensuring financial stability, but expressed uncertainty about \textit{how} these metrics impact them -- leading to a need for aggregate insights. D2 proposed collecting and analyzing \replace{different}{how} worker ratings and completion rates \del{for how they} affect the chances for getting more work\replace{.}{, while} W3 \replace{wondered about using}{wanted to leverage historical} work data to identify whether any past client reviewers were ``high value or high yield''---leading to subsequent work---as a basis for refining their work strategies (W3). \replace{To}{Lastly, to} support education around worker's financial risks, \add{worker} participants suggested insights on working conditions to support the creation of workplace health standards for gig work. D1 felt that if policymakers had access to worker data on the ``number of hours worked per day/week'', policymakers could ``establish regulations that can prevent us from working and ensure proper breaks  \dots  which helps reduce burnouts and health issues''. In addition, workers desire for a system that displays income, fees, and taxes, so as to give room for informed financial planning as well as support related education.

\subsubsection{Safety\add{: Overcoming Power Asymmetries and Physical Risks While Accounting for Worker Reports}}\label{safety}
\paragraph{\add{Policy Domain Experts' Concerns on Worker Safety}}
As non-employees, {gig} workers lack access to resources for ensuring safety while on the job\replace{. Because contractors of the ``unregulated 1099 economy  \dots  don't have a direct manager  \dots  [or] a person I can go to negotiate'',}{, } they ``don't even know where to go if there's a safety issue'' (P9). The lack of adequate worker provisions has not gone overlooked ``at the state DOT [Dept of Transportation] level'', where regulators like P2 are ``caring about licensing and crash history of people that are getting employed'' \replace{. P2 additionally {held} platforms {accountable} – } {and also challenging the accountability of platforms: }``What does that look like in terms of the companies protecting them from bad situations?''. \replace{The safety hazards that participants }{Policy domain experts} brought up \add{a wide variety of safety hazards} \replace{included}{including} the dangers of accidents on the road: caretakers staying in someone else's home, and women working overnight shifts. Since workers can provide a broad range of services, P9 notes how each type of work can entail ``very different safety concerns'', and there's a stark contrast between ``I'm worried my bike's gonna get hit'' and ``I'm worried about a passenger pulling a gun on me'' and ``when you're in someone else's house''. Socioeconomic factors further compound the risks since ``when you're in somebody's home, [there] is the social distance between the worker and the environment'' (P8). 

\del{\subsubsection{Safety}\label{safety}}%angie changed this 6-20-2024

\paragraph{\add{Worker Experiences of Safety Risks}} In each workshop, workers, \textit{especially those of marginalized genders}, expressed that work platforms lack concern for \add{physical} worker safety. Participants along specific intersections shared a distinct fear over the impacts of societal marginalization: is ``safety is going to be more compromised'' due to having ``ethnic name/non-white/queer/trans on your profile?'' (W3). \replace{Furthermore, workers}{Workers further} observed \replace{that the burden of safety is placed solely on the worker}{how the burdens of accounting for safety is shifted onto them (away from platforms/employers)}, increasing their vulnerability: ``There's a safety concern, and it feels like the onus of that safety is put entirely on the individual sitter  \dots  to ensure that you're safe'' (W2). \replace{\textbf{Worker safety is exacerbated by the power platforms and clients hold over workers being able to receive future work.} Platform metrics such as task cancellations and client ratings or reviews can impact a worker’s ability to receive new jobs and even continue working on the platform. This }{The fact that platforms and clients wield the power of future job opportunities over workers (via task cancellations, ratings/reviews, or slower new assignments)} can lead to \textbf{work coercion} where workers are compelled to accept unsafe working conditions to ensure good ratings for work and income stability. For example, workers felt compelled to continue working when ill\del{, fearing the adverse effects of cancellations on both current and future earnings}: W3 explained that a petcarer's profile can be negatively affected if they have to cancel multiple bookings, even if due to sickness. R4 shared that rideshare drivers feel obligated to accept rides with pets, even if they're allergic, due to the risk of deactivation, which offers no opportunity for rebuttal. F1 told us freelancers face pressures to uphold their reputation\replace{ for future work and financial stability, thereby submitting to jobs}{, causing them to bid to jobs} with terms and conditions that encourage overwork.

Workers also shared a prominent concern around physical safety imposed by \textbf{\add{uneven power relations with} clients}. Since platforms prioritize clients over workers for revenue, clients can easily exploit the power imbalance to harm worker safety. 
According to W3, workers are subjected to background checks but \emph{clients are not,} allowing them to create new profiles:
% In contrast to the protocols put in place for clients' safety, W3 vividly recounted an instance where:

\begin{quote}

``A friend who does services in the metro Atlanta area [had an experience] where there was an owner who had \dots undisclosed cameras and he was asking specifically young women to do house sits. And so, she reported this and obviously was allowed to terminate the sit \dots Later, she noticed that the person created a new account using the long version of his name, and she also saw that same dog under a different client.'' 
    
\end{quote}

Finally, power differential contributes to workers being pressured to accept invisible work outside of their original assignment’s scope. W2 explained petsitting operates in a legal ``gray area'' so ``there's a lot of invisible work and no ability to delineate scope of work''. W3 added reading about cases where ``people are being asked to do stuff like clean houses or provide childcare or do other things while they're being only booked on this platform for pet sittings.'' 

\paragraph{Data needs to promote \del{physical }safety:}\label{h.xevzq1wr781s}
\add{Policy domain stakeholders} P2 and P9 were interested in \replace{how data can}{using data to} improve transparency on current safety standards, incidents, and concerns workers have, as well as set future standards of safety for gig work. P2 explained that, ``delivery driving [is] a job that does not get the credit for being as dangerous as it is,'' and wanted to survey workers on how safe they feel doing their jobs. From workers, P9 wanted to understand the \textbf{experiences of workers who have felt unsafe} as well as ``were they able to be resolved? Did [the worker] know of a resource to go to?'' From the platform, P2 wanted data about \textbf{safety incidents} that have jeopardized workers and more transparency around platforms' policies and trainings in place to protect workers. P2 pointed out a desire to set safety standards, including procedures for reporting incidents and support measures to help affected workers return to work.

\del{\paragraph{How Data Can Support \& Provide Insights on Safety.}} Workers wanted information to not only understand what and how frequently safety incidents are occurring, but also to empower them in safety-oriented decision-making. To achieve this, they suggested combining \textbf{safety data, reports} from gig workers on safety incidents, and other types of gig work \textbf{data specific to their geographic region} to generate relevant insights. For example, R4 suggested gathering reports of carjackings and associated locations from rideshare drivers to support awareness of other drivers\replace{. R4 also suggested}{, as well as} analyzing patterns of drivers declining trip requests by neighborhood in conjunction with the neighborhood safety metrics to draw attention to regions drivers’ feel unsafe working in. \del{Interestingly, he mentioned a concern about passengers being discriminated against and how the analysis of driver decline or cancellation rates by geographic area could ``shine light on the discriminatory practices''.} To inform their decision-making, workers wanted ``learn about how other gig workers navigate in less secure areas'' (D4)---with D4 and R3 suggesting this could be done through sharing with one another experiences of safety at different drop-off locations. Though they will still feel pressure to accept certain trips they are wary about, having this qualitative, experiential data can help workers feel more prepared. 

\del{\section{Findings from the Interviews of Policy Domain Experts}} \label{pde}
\add{\subsection{Stakeholder-Specific Initiatives: Stress \& Experience-Sharing} \label{specific}} 
% \textcolor{blue}{Summary of unique initiatives?}
\del{Below we present four core initiatives the policy domain experts mentioned in interviews{, each followed by} concrete data that participants thought can support the initiative. }
\del{We summarize {challenges policy domain experts expressed around ownership, privacy, trust and accommodating diverse worker types.} {While we spoke to participants from varying levels of government, substantial similarities surfaced in the ideas and initiatives discussed between different levels of office. Hence we do not distinguish between local- and national-level insights in our findings.}}

\del{\subsection{What do policymakers care about, and how can data support them?}}\label{h.ohekeel96pzt}
\del{Policy domain experts emphasized various prioritized initiatives – which included \textbf{expanding equitable access, approaching fairer pay practices}, as well as \textbf{reducing (care)work-induced stress}. They expressed excitement about worker-based data sharing, describing how worker statistics can help advance related policies. For example, policymaker participants largely expressed interest in viewing \textbf{aggregated data over raw data,} since the former is better suited for informing or funding resources and program development for workers (P1, P6, P7, P11) and is also less prone to manipulation by ill-intentioned actors (P5). In addition to descriptive statistics, participants also wanted to see \textbf{qualitative data} to help them understand job quality and workers' stories, such as P7 asking, ``what happens after they get here?'' (P1, P7, P8, P9). }

% emphasize reorg of findings in summary of changes, taking suggestions from reviewers -- instead of having two sections separately reporting findings (original structure), we completely restructured findings

\subsubsection{\add{Policy Domain Experts' Concerns about} Stress}\label{stress}
\replace{Participants}{Policy domain experts} identified a variety of factors that pile onto gig workers as stressors: the need to navigate between different roles, a lack of temporal stability, as well as the shortage of financial and mental resources. In particular, participants expressed concern around the stress imposed on gig workers with caretaking responsibilities, including 1) those who give care to their own family members or 2) workers in caretaking gigs.

The \textbf{high variability of schedules} in gig work contributes to financial and mental strain\replace{. The touted flexibility of gig work trades off with}{, trading off with flexibility and} agency since workers' earnings depend critically on consumer demand (P7, P8, P9) \replace{, thereby causing work precarity and forcing}{--- this forces} workers to accept most gigs during times of low demand, regardless of personal constraints. P8 empathized with ``already time-starved'' workers, who are then further deprived of ``time [and] distance [when] spent driving idly and unpaid'' because platforms pay workers for the effort of the gig itself, but not the effort spent searching for gigs (P9). 
To overcome the work precarity\del{ that is inherent to gig work}, some might take on a \textbf{multitude of roles} to meet financial needs\replace{. However, workers}{, but workers} serving multiple clients \replace{bear the risk of meeting especially}{can then encounter} stressful situations. ``If you're working two different jobs with two different employers and there's a conflict, you have to figure it out, [and] that is very stressful for people. Plus, it's kind of an indication that the job you're doing is inadequate in terms of living wage.'' (P8).

Caregiving is a stressful and often-overlooked form of work, causing many caretakers to leave the industry in favor of higher-paying jobs. P1 illustrates \replace{the}{how \textbf{caregivers receive}} \textbf{meager pay} \del{that caregivers receive}\textbf{for stressful tasks} ---``if a childcare worker can make \$15 an hour, but they're spending all day \dots changing diapers and dealing with the emotions of children and kids screaming'' then it's no wonder that we have a shortage of childcare services.'' Participants also expressed concern for workers with caregiving duties within their own families\replace{. Currently, the nation}{, especially since the nation currently} lacks systemic support for working parents. P1 pointed out ``the high cost of childcare that isn't subsidized by the government'', which is a result of ``the federal requirements for just employment  \dots  [being] not evolved in that space when it comes to \dots parental benefits'' (P2). Consequently, the lack of ``access to childcare for moms'' (P2) ``is forcing more and more women to go back into the home and leave the workforce'' (P1).
% To mitigate the shortage of childcare services, P1's organization is ``advocating for an industry shift in terms of what pay structure for childcare and caregivers looks like  \dots  [to] stop the bleed (so to speak) of childcare workers from leaving that industry'' (P1).
% , because the current inadequate supply of affordable childcare underscores the ``systematic structural issues that are prohibiting women from building economic stability when they are preparing for their family'' (P1). 

% Participants paid particular attention to \textbf{workers with caregiving responsibilities,} which included both 1) gig workers with caregiving responsibilities in their own family and 2) workers in caretaking work positions. On the one hand, caregiving is a stressful and often-overlooked form of work, causing many to leave the industry. However, the shortened supply of caregivers in turn limits parents' access to childcare, which pushes working mothers back into domestic roles. 

% As an example, ``if parents are paying \$16,000 a year for childcare, but childcare workers are only getting \$15 an hour, they're falling off'' (P1). 

%\textbf{Data needs to understand and support}\label{h.ll2ktazif6ci}
\paragraph{Data needs to understand and support \replace{caretakers}{stressed workers}:}  

Some participants wanted data to understand longevity and quality of life for workers, though P9 recognized \replace{this as a challenge that's}{that such data} ``\del{something we }don't quantify very well''. P1 and P7 \replace{wanted}{sought} accounts of why workers began gig work, if they view it as ``a stepping stone'' or ``something they see themselves doing kind of indefinitely'' (P1), and ``what happens to them after they get here'' (P7). P8 was curious to use data to assess job quality and living conditions, ``I'm always thinking about, can I live like this?''
% Add sentence about multiple roles
To \textbf{quantify work precarity}, participants wanted to investigate \replace{what factors}{factors that} lead to fluctuations in worker wages \replace{. Some}{--- some} suggested quantitative measures \del{to analyze this} --- whether seasonal or holiday fluctuations affect the profitability and stability of gig work (P6), \replace{what the likelihood}{likelihoods} of receiving tips is, and \replace{what proportion}{proportions} of a worker's income \add{that} comes from tips (P3, P4). 

To support caregivers, participants wanted to use data to understand the toll of gig work on mental well-being. P1 and P5 \replace{wanted to quantify}{imagined quantifying} the precarity a worker experiences with their stress levels by assessing the consistency of workers' schedules, hours worked, and {the availability of} work. P9 explained that for jobs like caretakers or cleaners, insufficient notification could leave workers scrambling to find childcare in time to commute to a client's location. P2 and P3 \replace{wanted data on}{desired to know} whether workers have access to work benefits like childcare and health{care}: % , P3 advocating for at least basic healthcare benefits for gig workers
``all should have access to even \dots minimal benefits \dots so people can get healthcare and remain healthy and support a family''. P1 \replace{, interested}{explained her interest} in the correlation of stress on the tenure of a job for working mothers or childcare workers\del{, explained}, ``we do find that information about the experience on the job, or the amount of stress, directly correlates again to the opportunities that women, and in specific, working moms, have in terms of choosing to do certain types of work or not.'' To that end, P1 wanted \textbf{both qualitative and quantitative data} to analyze whether people ``stay in certain gig work or move on to something else that is of equal pay and potentially lower stress and more possibility for future growth.''

%\subsubsection{Other initiatives and work assistance programs}\label{h.v3px1c1ygbdo}
%Participants reported various other initiatives of concern as well as existing programs that they support or had worked on. For instance, P5, P7 and P11 raised the issue of how gig workers' limited rights to unionize are shortcutting them of many needed benefits and protections. Health risks were also of concern to P9 and P7, who called for OSHA protections and workers' compensation; meanwhile  P1 and P3 both championed for paid sick leave. Education was also emphasized to help workers gain financial literacy and provide them skill sets for acquiring gigs, and guide returning citizens back to the workforce. Finally, participants also described various programs or initiatives they supported or worked on, which included a local co-op for gig workers, a program for helping returning citizens find employment, as well as a self-sufficiency that strives to guide workers toward higher earning employment. However, since it is currently unclear how worker-contributed data can help advance such initiatives, we include in the appendix participants' detailed descriptions of these proposed initiatives and programs.

\subsubsection{\replace{Sharing}{Worker Desires to Exchange} Strategies, Experiences and Context for Advocacy}\label{h.pplzn44x8y2i}
Workers often asked about and shared experiences and understanding of platform features and functions during sessions --- such exchanges may also be facilitated by a data-sharing system. 
Generally, workers exhibited a desire to learn from others' experiences and strategies. Driver R2 was keen in ``see[ing] if there are some people who are experiencing quite the same [as] what you're experiencing and how  \dots  to better deal with some stuff that come[s] up along.'' Through sharing, workers can teach each other work-related knowledge such as ``learning about fuel consumption of a certain vehicle \dots you might decide to change to that more efficient vehicle'' (D2). Meanwhile, D1 sought advice on how to deal with being busy during peak hours: ``How do they navigate when it's peak times  \dots  and they're not available?'' For petsitting, W2 sought wisdom on how taking discounts impact earnings: ``from an overall community perspective, understanding what happens when you discount would be really beneficial''. Petcare and freelance pay rates can vary based on the quality and types of services, leading workers to seek insights from others regarding \textbf{standard rates and pricing strategies}. W2 recounts looking up ``what other people are doing in terms of how to set your prices competitively'' so they can emulate them. Freelancer F2 hoped to learn standard rates from more senior workers: ``if a senior is charging maybe some kind of fair amount, then you want [use that] to determine the amount you charge'' (F2).  

Additionally, there is a noticeable gap between the knowledge of many workers and available protections: workers may encounter situations that seem to ``violate some kind of worker [right, i.e.] not having access to a restroom in your workplace \dots there's no education on that,'' leaving them in unfair positions of not knowing what to do (W3).
Workers \replace{want}{also saw value in} training programs \replace{covering}{that cover} these areas\replace{ as well as an understanding of how such education impacts their earnings and work satisfaction }{, in addition an understanding of how such education would impact work earnings and satisfaction} (D1).
% Second, workers also desire formal rights and protections against platform exploitation. 
Beyond education, workers call for advocacy contextualized by gig work to advance rights and protections, as policymakers fall so far removed from gig labor that they ``don't even know what to ask'' to support worker-centric initiatives (R4). At the moment, there is ``zero responsibility [on the platform-side] \dots they do what's necessary by law and the rest of that you're just on your own'' (W2). Without strong or accessible resources to uplift them, workers are subject to disadvantageous positions in favor of platforms' bottom-line (R4). Under these conditions, workers wish to provide policymakers context on unique struggles (e.g., unfair wages, violations of privacy, lack of fiscal knowledge, burnout, health issues) that incur physical, mental, and financial harm (R4, W3, D2).

\paragraph{How data can support sharing of strategies, experiences and context: }Interviewees indicated broad interest in accessing other workers' experiences and ways to learn the ropes of a gig job quickly and build a profile for themselves. For this, workers desired to \textbf{learn working strategies} such as specific delivery routes to know how others navigate insecure areas (D1, D4), ``how other workers handle difficult situations [or] how they take their orders'' (D1), information on ``traffic conditions, or road closures or any other factors that can impact my travel time'', which would even enable them to practice ``an alternative workflow'' where ``by taking that data and seeing where everybody's working, you don't go there  \dots  [which] can be a competitive advantage'' (R4). To understand expectations, W1 and W3 wanted to know projected commute times for certain neighborhoods while F1 wanted geographic locations to understand the cost of living standards. Combining such geographically-specific information with the
\textbf{rates others charged for services} (F3, W3) and ``what other people  \dots  would charge extra for  \dots  and \dots would consider the standard'' (W2), can help workers set their own standards for service rates.

% Incorporate quotes in other sections, such as fair pay or 5.2
% \subsubsection{Advocacy \& Education for Worker Protections}\label{h.rrgldkacru1f}
% %angie edited 6-21-2024
% Workers strongly aligned on the need for advocacy and organized efforts to champion and support their pressing concerns. Participants recognize that ``it's really hard for [workers] to individually advocate for [their rights]'' especially when ``the people that are accepting [gig work] are in relatively financially needy situations and really want to take whatever work they're being offered'' (W3). Participants convened on two critical areas where advocacy and worker education is essential: \textbf{financial literacy and transparency} as well as \textbf{policies and regulation for worker protections}.

% content moved to other sections

% \paragraph{How Data Can Support \& Provide Insights on Advocacy and Education.} 

Workers' suggestions of insights and data related to the prior sections also apply here for creating educational programs and  advancing policy or regulation around worker health and safety. For instance, workers’ suggestions for data around metrics and stories about workers’ safety and discrimination experiences (See Section \ref{safety}) could be shared with policymakers for crafting formal rights or protections for gig workers against platforms. Access to this large scale dataset would provide insights into the health and wellness of workers, and the amount of workload they are responsible for, key cornerstones in asserting worker causes.

\subsection{\replace{What challenges do policymakers perceive for a worker data-sharing system?}{Practical Considerations: Worker Diversity, Privacy \& Trust Perceptions, Ownership \& Access}}\label{h.tycl8ab0p8ie}

\add{When prompted, participants raised several practical concerns they can foresee around worker datasharing systems. Policy domain experts worried about how the term ``gig work'' is too broad to accurately capture the diverse experiences of workers on the ground. For workers, such diversity of task domains and workflows produced divergent preferences on \textit{how} to best upload data. While privacy preferences were never discussed in full detail, both stakeholder groups emphasized the importance of establishing trust by providing (1) clarity around what privacy policies are, (2) effective communication when there are changes (3) ways of opting out, as well as (4) transparency around who gets to access certain data. Relatedly, worker participants offered perspectives on which stakeholder groups (i.e., peers, policy domain experts), while policy domain experts discussed potential institutions/organizations who might be good candidate for owning the data(-sharing system).}

\subsubsection{\del{Accommodating }Diverse Worker Types \add{Lead to Heterogeneous Preferences for Methods of Data-Sharing}}\label{diverse_types}
P\add{olicymaker p}articipants explained \replace{that}{how} the language around gig work varies significantly between researchers, policy experts, and the public. P9 stated \replace{how}{that} the general public's perception of gigs consists of platform-based work but not childcare or intermittent work, which is seen as ``under the table'' work, whereas the Fed considers gig work to be all forms – job to job, temporary work, tutoring – not just platform-based gig work. Inconsistencies between the perception around different gig work task domains can hamper grassroots activism, and the catch-all term ``gig work'' runs the risk of \textbf{generalizing nuances} between platforms. \replace{In P9's conversations with workers, they}{P9} learned to be explicit about task types \add{when talking to gig workers, who often do not use the term to self-describe}: ``How would you talk about work like TaskRabbit or DoorDash? We [gig workers] would say TaskRabbit and DoorDash.'' Certain platforms give way to worker cultures with differing financial obligations. Due to differences in ``zones of city employment'' (P7), even workers on the same platform can experience financial differences. Yet, insights can be drawn by comparing platforms from various angles. While there is a ``different exchange of resources'' (P8) between caregivers and deliverers, \replace{caregivers are algorithmically}{they are algorithmically} managed much like rideshare drivers (P7). Caregivers and petsitting – both considered care-based gigs – have drastically different levels of service variation (P7) and legal ramifications (P9). Meanwhile, asset-based platforms like Airbnb tend to contribute toward supplementary income, whereas ``people who did like TaskRabbit or Uber were using it as primary income'' (P9). When compared to physical gigs, remote freelancers face less safety risk but are more at risk for theft of services (P9).

\del{\subsubsection{Preferred Methods of Contribution}\label{h.ma4hocgxchdb}}
\replace{Participants also discussed}{The different types of tasks gig workers engaged in affected their} preferences on \replace{how often they want their data, such as job request time and pick-up location,}{what gets shared, how it is submitted, and how often it is} to be uploaded to the platform. 
We provided workers with examples of data formats, e.g., app screenshots, CSV files, and automatic data point connections. 
One popular preference \replace{by workers from all four task types is to have the}{that workers across domains shared was the idea of} \replace{data automatically collected and uploaded}{automatically collecting and uploading data} through an app for convenience and reduce the strains of manual upload (W2\replace{, W}{-}3, D1\replace{, D2, D}{-}3, F1\replace{, F}{-}2, R4)\replace{. A rideshare driver, R4, raised a consideration}{, although R4 raised the concern} that manually uploading data could impact data quality if drivers ``cherry pick [their] good weeks and not [their] bad weeks or vice versa''. On the other hand, some workers prefer to manually upload their data on a sparser schedule due to security concerns (F2). Others preferred manual upload due to job specificities. As a freelancer, F3 was accustomed to updating their clients daily to set expectations and receive feedback, so \replace{they wanted}{thus preferred} daily data uploads. As a pet sitter, W2 preferred a monthly schedule as there is ``some seasonality'' to the job and \replace{having the data collected in real time would be}{this real-time data collection would cause} ``a drain on [their] account and phone''. 

Workers also explained their \textbf{preferred device for data uploads}. Most \del{participants }wanted to use their phone and computer to upload data formats of texts and emails for convenience (R1, R2, R4) or ease of customization (F2, F3). D1 proposed an offline app as an alternative to a website as it ``allows you to share data without getting connected to the internet''. R4 preferred using CSV files, but was unsure if other drivers are tech savvy enough to follow the same process, suggesting app screenshots can make for a better user experience for other drivers. 

\subsubsection{Privacy and Trustworthiness}\label{h.63yxjp6nz5ae}
\replace{Privacy of the data shared on the system is less of a concern for participants. }{When prompted, \uline{policy domain experts} exhibited minor to no concerns around privacy violations.}
% Particularly, P3 wanted to have a clear plan on ``how the data sharing would be facilitated, what is envisioned, or how it would be accessed or what it would look like''. 
Given the information will be presented at an aggregate level, P3 ``would not be concerned about [reveals of] identifying information''. The only exception to this is when the system wants to apply privacy-preserving techniques like k-anonymity to ``a smaller specific company and the identifying information was obvious because there are so few workers.'' Similarly, P5 believed that ``the safeguards put in place by the academic institutions to be able to do surveys are pretty good generally,'' so there are no specific concerns about privacy as long as \del{the system obtains the}necessary consent \add{is obtained} from the participating gig workers. P5 further clarified that the policymakers should only have access to the system data through ``packaged white paper'' and not the raw data. On the other hand, there are some concerns over the workers' trust in the system, depending on who owns the data. Particularly, P9 questioned ``do workers feel differently about providing their data to the Department of Labor?'' because their interactions with workers revealed that ``sometimes [workers are] uncomfortable \dots about being in a conversation with the Federal Reserve.''

\del{
\subsubsection{Privacy and Trust}\label{findings-privacy}}
%grabbed this and edited it from commented out section before on privacy
\replace{In addition to privacy preferences for sharing data with others, participants also reflected on}{In our conversations with \ul{workers} around privacy, they expressed} the need to
understand a data-sharing system's policies around privacy and data ownership before engaging with it. \replace{Participants}{Workers} often explained establishing trust and a system's privacy policies going hand in hand (R3, F2): ``You have to trust them with your data to ensure that [they] would keep it private'' (F2). People worried about data being shared (R4, D2), sold (R1), or leaked (F2, W2, D1) to nefarious actors who would misuse it. W1 and W2 were specifically nervous over location data, especially if released to past problematic clients. Participants also wanted to ensure they maintained full ownership over their data, including the ability to revoke data access should they change their mind (D2). Relatedly, W3 wanted a data-sharing system to avoid scope creep---continually changing terms and opt-out conditions---which would burden the worker to regularly review terms and conditions and learn how to manually opt out of new data collection: ``A lot of times [now] is you're discovering that the latest terms and conditions you had to accept automatically opted you into data collection for AI and then you have to manually go and figure out how to get out of it.''

\subsubsection{\replace{Ownership}{Stakeholders Who Should Share Ownership \& Access Data}}\label{ownership}
\del{When envisioning a worker data-exchange platform, participants discussed in depth the question of ownership. }
Currently, there is no consensus \add{among \uline{policymaker participants}} on which entity should have control over collected data\replace{, and below, we}{. Instead, we} outline \add{below their} rationales of support for and against particular groups as owners.
%For instance, P1 considered it appropriate that owners include ``the people who are providing their data'', while P6 thought it is necessary to clearly communicate to workers who holds responsibility for the data platform: ``the biggest thing, once it's developed, would be to get [the message] to the gig workers [on] who's responsible for it''. For P3, having the data-exchange platform would directly benefit the workers as it ``creates a paper trail and notes and writes down every infraction'', enabling them to document transgressions and thereby fight for their rights. P3 imagined workers using the platform as ``the inverse of an employer who is seeking to fire an employee'', so as to provide an environment where ``the worker can note that they felt that pay was withheld or discriminated against'', so ``they [would] have a case built to be able to say ‘Look: we warned you, we warned you, we warned you now we're gonna fire you.''' 
%create it initially: ``how would that get launched? How would that get lifted up and become a reality for them?'' (P4). Maintenance poses another issue – though P5's ``default [option] would be the workers that own it'', they worry that such ``presents different organizational capacity issues'' since workers have no dedicated concrete funding for the upkeep of the platform and it would be presumptuous to assume ``that workers have the time and energy to manage the data'' (P5). P9 and P4 also foresee data management as a challenge requiring staff time and expertise to handle.

\begin{itemize}
    \item Many supported the idea of giving \textbf{gig workers} back the control over their own data (P1, P3-7). However, P4 and P5 worried that workers may not have the bandwidth to develop and maintain the platform. 
\item Some participants (P1, P4-6) suggested \textbf{researchers/universities} as candidates for owning the data platform. But P4 raised the concern that universities might not ``maintain this [the research project] on an ongoing basis,'' whereas the data platform would need a permanent home, plus there are expected scaling difficulties because ``one research institution \dots may have a lot of trust where [they] are and may not in [different locale]'' (P9).  
\item Participants (P1, P4-7, P9) also brought up a variety of \textbf{(labor) advocacy organizations} as potential owners. P9 recommended ``intermediary labor organizations or groups that have trust that would make workers less reticent to share their information'' as potential owners while allowing the Bureau of Labor and Statistics to access ``regular data on gig workers'' as ``many institutions that do research on labor  \dots are extremely reliant on [the BLS].''
Advocate P6 would want to share {such a data system} with their leadership team to ``put in [their] channels with partners and with community members.'' However, entirely handing over management and governance to advocacy groups runs the risk of political biases, since ``advocates are generally very subjective \dots I would be concerned about more biases \dots or the way that the data is aggregated.''
\item Most participants (P1, P2, P5, P6) resisted the involvement of \textbf{platforms and employers} in providing and managing data for policy-making purposes.
%, a couple advocated for their inclusion to ensure a balanced decision making process. 
According to P2, asking platforms to directly provide the data would result in ``a rosy picture that is inaccurate.'' 

\end{itemize} 

Instead of governance and ownership by a singular stakeholder group, participants suggested the alternative of a \textbf{shared ownership between multiple stakeholders} (e.g., workers, advocates, government agencies, and neutral third-party organizations). P8 believed that the data-sharing system {should strike a} ``balance between workers, designers, policymakers, employers \dots to weigh in on this because their perspectives are different, but also each can learn from the other.'' 
Similarly, P7 wanted advocate-worker hybrid governance, where advocates would provide infrastructure, but the data-sharing system is controlled by ``worker governed entity  \dots structured as  \dots  Limited Liability Company (LLC)'' where ``a lot of seats [are] given to gig workers.'' 
\newline

\add{While \uline{worker participants} did not actively express interests in self-owning the datasharing system, they did hold opinions on whom they prefer to share data with. We summarize below the stakeholder groups they might consider sharing with, the types of data to share with each, as well as rationales and motives for allowing them such access.}

\paragraph{Peers}
Most participants approved of sharing aggregate data but not individual data with peers, primarily due to concerns related to competition. D1 described ``if my peers can have access to the same data and insights that I have, they may target the same high demand opportunities that I normally rely on. And things will lead to reduce the amount of my earnings.'' Those willing to share individual data expressed a sentiment to help others learn (``I'm learning from them, they're learning from me'' --R2); acceptance if data sharing is an equal exchange (F2); approval for sharing data with those from different cities, preferences or work patterns (W2, D2); or had no specific concerns preventing them from sharing de-identified individual data (W2, F1). D1 provided an interesting perspective of willingness to share her data in exchange to view others' as a way to motivate herself: ``If you are the best, and I want to be like you, I have to push myself. So it's like a motivation or a challenge.''

%Those wishing to block individual data from other workers explained having concerns of data privacy breaches (D1, D2) or losing their competitive edge by having their strategies copied (D1, W3): ``I personally don't see myself wanting to model my business practices exactly on one specific other person, and I would feel kind of weirded out about somebody else doing the same for me.'' (W3). A couple made exceptions for sharing individual data only if it is with those from different cities or different preferences or work patterns (W2, D2). And while most were willing to share group level data with all other workers --- ``I don't mind about group level because at least we're all sharing the same information, not same, but it's just the experience'' (W1) --- D1 expressed wanting to block all data out of an abundance of caution for protecting his competitive edge.

\paragraph{Policy domain experts}
R4 wanted policymakers to have access to individual and aggregate data to properly investigate worker concerns: ``I think policymakers should have an idea of how egregious the rates \dots the fees that Uber charges.'' However, most others preferred to only allow policymakers to use aggregate, group-level data, due to concerns of exploitation (D2, W3, D1), or government micromanagement (F1). They viewed aggregate data as sufficient for specific initiatives they wanted policymakers to focus on, including wage theft (F3, D4), worker safety or worker and passenger discrimination patterns (F3, F4), and equitable job allocation (W2). D2 and F2 emphasized sharing data disaggregated by demographics with policymakers to center inclusivity: ``that would help in them making a collective decision \dots they come up with a solution for everyone'' (F2). W2 shared one hesitancy about how to ensure data integrity, suggesting that off-app transactions can skew data patterns: ``You're not necessarily getting the full picture. Someone might Venmo me a tip \dots how would you collect that data and how would that affect how it looks?''

\paragraph{Other stakeholders}
A few participants also mentioned sharing varying levels of data with customers, family, and even lawyers for awareness purposes. D2 and R4 wanted customers to see data related to platform tactics: D2 described being on the receiving end of customer complaints over high prices and fees, and giving them access to data to understand charges could ameliorate this. R4 suggested that sharing data with riders on ``what we get paid versus what they pay'' could advance efforts for platform transparency by ``creat[ing] an uproar from the customer side of the house'' to pressure platforms. Interestingly, F2 and D2 both suggested allowing gig platforms to access their data, D2 elaborating it would be necessary for ``assigning accounts or doing the maintenance of the app or website''. 

\del{\subsection{Preferences and Concerns for Data-Sharing}}
%angie working thru this 6-21-2024

\section{Discussion}\label{h.yin5uhe4br0e}
\anedit{In our study, we learned about priorities workers and policy experts share, separate goals they emphasized, and their ideas for how a collective data tool can support these. Based on our findings, we first share design implications for data-sharing systems that enable worker and policymaker alignment on worker initiatives. Then we reflect on practical challenges and considerations (e.g., privacy, trust, ownership and transparency) for creating a data-sharing system.}

\subsection{\anedit{Designing Datasharing Tools for Common and Distinct Priorities}} \label{discussion-multi-stakeholder}

\subsubsection{\anedit{Supporting Shared Initiatives: Identify Data Types, Build Public Awareness \& Affect Policy}}
%implications of how to design for shared initiatives

\anedit{Our results indicate that workers and policy domain experts share common ground on the following set of worker issues: \textit{Discrimination \& Equity}, \textit{Fair Pay} and \textit{Safety}. This reciprocated interest suggests that for these issues, 
%instead of needing to \textit{persuade} these stakeholders about the relevance of such issues, 
worker-centered tools can be developed to 1) enhance multi-stakeholder investigations into what data (types) is most important to collect, 2) strengthen awareness-building and educational programs, and 3) help draft language for policies and standards. %We encourage researchers exploring data-sharing systems to begin by building on this common ground. 
%In particular, shared priorities can be bolstered through features that 1) elevate workers' expertise from lived experiences and 2) highlight policy experts' expertise around policy processes and existing aid programs. 
We offer recommendations of how to practically approach such objectives
for the three initiatives participants aligned on.}
%To advance common goals, we suggest ways that researchers and developers creating data tools can enable bi-directional knowledge sharing amongst different stakeholders advancing worker well-being.}

\paragraph{\textbf{\anedit{Leveraging Qualitative Data to Pinpoint Drivers of \textit{Discrimination \& Inequities.}}}}
\anedit{
% Workers and policy experts aligned on the importance of probing for \textit{Discrimination \& Inequities}. 
While both stakeholder groups described the need for traditional demographic data to investigate occurrences (e.g., race/gender pay equity), workers also emphasized the potential of non-traditional factors for signalling risks for inequitable treatment and discrimination: W3 and W5 worried about the impact of ``visibly ethnic names'' on earnings, while D2 pointed out how immigrant delivery workers must resort to more risky modes of transport (e.g., biking), in lieu of obtaining driver's licenses. %Additionally, workers we spoke to strongly emphasized the value of hearing each other's anecdotal experiences with discrimination.
%This aligns closely with how workers and policy domain experts talked about using data to understand equity and discrimination---gig workers we spoke to emphasized the value of hearing each others' anecdotal experiences with discrimination.  
%
Such qualitative on-the-job experiences can generate novel metrics and critical (but previously overlooked) factors for policymakers to enhance their understanding of discrimination and inequities. So that when they draft the policies that \citet{van2023migration} calls for at the intersection of immigration and employment regulations, they can accurately account for the often latent experiences of workers at the margins. 

To strengthen efforts against discrimination and inequities, researchers can focus on tools that facilitate collaboration between policymakers and workers in identifying \textbf{key attributes of their work to track}. 
For instance, in response to worker's desires to hear about each others' personal anecdotes (\ref{data_disc}), one design could implement a \textbf{multi-stakeholder-facing interface to enable experiential reports} of discrimination by workers to policy experts.
On the workers' end, individuals may record qualitative narratives of their experiences, and additionally create tags for (parts of) posts to signal unexpected, biased and alarming aspects of their work that may serve as potential measurements of discrimination. 
On the side of policy experts, narratives/stories can then be surfaced and grouped by tags, or even sorted based on preferences expressed by worker groups (via voting through mechanisms such as likes or upvotes), so as to help policymakers identify concrete experiential evidence that reflect workers' or their own priorities.
% Additionally, aligned with D2's suggestion to show policymakers disaggregated statistics, worker-created tags could be presented as new quantitative data features for policymakers to consider through recommended, experimental visualizations or tables based on the tags.% such as spoken language capabilities as D1 suggested) could be displayed on these dashboards (even in a speculative way)
}

\paragraph{\textbf{\anedit{Educating \& Raising Awareness on Factors that Impact \textit{Fair Pay}}.}}
\anedit{While the idea of identifying and aligning pertinent data to reveal discrimination patterns can also apply to \textit{Fair Pay}, policy experts and workers also indicated interest in using data to support educational programs on fair pay to help workers understand if and how they can earn a livable wage doing gig work. %, explaining platform tactics and specialized terminology to help us understand their concerns around earning a livable wage. 
For instance, both W3 and P1 (and prior works \cite{xAXX, finlit}) viewed financial data as valuable for providing immediate benefit to workers (for understanding risks and whether livable wage conditions are met) and policy experts (to support drafting and passing of legislation) while minimizing time and effort required to collect data at scale.
%Finally, workers vehemently call for policymakers to provide accessible education about the \textbf{financial risks and responsibilities of gig professions} \replace{and provides clear \textbf{guidelines on worker boundaries and rights}. Currently,}{since }there is \add{currently} a significant lack of financial \replace{knowledge}{understanding} amongst gig workers around income structure, taxation, and metrics they should track and monitor\del{for financial well-being}. This knowledge gap enables platform-side exploitation, as ``a lot of gig work in general prey on people not being financially educated or not being able to forecast what is my actual earning going to be from this? How am I going to set aside money for taxes on all of this?'' (W3).  w3, p1
%
Thus, work tracking systems that ask for data contributions from workers should go beyond collecting data around expenses and time-tracking to also help them directly answer questions around meeting basic financial needs such as ``Am I making enough money?''. This can be achieved through both \textbf{visualizations of statistics} and descriptive overviews of financial data --- examples include personalized and straightforward summaries (on a task-by-task basis or across timespans) about how much they are netting, detailed but \textbf{digestible breakdowns of costs and earnings}, or estimations of gross earnings based on similar historical instances.
% , and potentially leveraging existing worker data tools like data probes by \citet{WRX9}. 
Such features should be made available to workers regardless of whether they partake in training. However, their value might be enhanced for some workers if paired with training workshops on topics like financial and algorithmic literacy, as well as gig work risks more broadly (\ref{pde_pay}).
%a system should explicitly remind workers about hidden costs and invisible work. For example, when workers share their data, a system can also nudge workers to upload expenses information by displaying exhaustive expense categories so workers do not overlook efforts like sunk costs. Or a system can pose questions related to invisible work, such as asking care workers to document time spent transporting to a client's home before a task began.
%

Workers also highlighted the knowledge gap that riders and policymakers have about how basic platform operations, suggesting the need for educational programs or tools that can (1) \textbf{inform \textit{policy experts}} about how underlying algorithmic practices (e.g., dynamic/upfront pricing) perpetuate \textit{Fair Pay} issues that workers experience such as subminimal compensation (2) \textbf{alert the public at large} about such phenomenon. 
Towards the first point, a data-exchange platform can gather worker anecdotes and data that serve as content (or at least resources) for such educational programs, in line with \citet{policy_probes}'s findings on how worker data probes can assist with educational efforts for policymakers. 
Towards the second, data can be used to raise public awareness as a spur for policy or regulation creation. 
    For example, as D2 and F3 suggested,  visualizations and aggregations of historical data (possibly through interactive tools) can help show the public the extent to which platforms overcharge customers and undercut workers, thereby raising awareness and public outcry that can motivate policy advancements around fair pay and platform transparency. 
% (what was that example haiyi gave ages ago about some government agency and the airlines??)}
}

\paragraph{\textbf{\anedit{Exposing Power Asymmetries that Threaten Worker \textit{Safety}}.}}
\anedit{Participants (of both stakeholder groups) pointed out the \textbf{power differentials of platforms and clients over workers}. For example, a poor client rating can keep workers from getting future jobs, while platforms can limit work opportunities through slower new assignments or even deactivations. The combination of pressures from higher-power actors often forces workers to accept jobs despite unsafe or unfair conditions. Corroborating prior work \cite{5qBZ, beyond, sannon2022privacy}, we observe client harassment as an additional relational factor where that puts female workers at higher risk (\ref{safety}). 
% Policy experts specifically expressed a desire for data---especially experiential (P9)---to support their efforts advocating for safety standards. 
% Workers desired qualitative data too as this can help them feel more informed when they have no choice but to accept work tasks to continue earning an income (). They also pointed out the significance of quantitative metrics, such as client ratings of workers as an element of control clients and platforms have over workers.
%
Unlike other discussed initiatives, policy experts indicated a direct link between \textit{Safety} and the use of data towards creating worker-centered labor and safety standards. This desire aligns with existing efforts to establish clear-cut modes of recourse for gig workers who have had their employment suspended due to client accusations. However, such mandated standards are currently only pursued in a handful of states (e.g., Washington State \footnote{https://lni.wa.gov/workers-rights/industry-specific-requirements/transportation-network-company-drivers-rights/resource-center-and-deactivations}, Colorado \footnote{https://leg.colorado.gov/bills/sb24-075}) and are usually specific to rideshare drivers. 
% In the case of Colorado, the recent legislation requires rideshare companies to create and disclose a deactivation/suspension policy to drivers, and compliance is so far only required ``on and after June 1, 2025''.
% , and comply with it ``on and after June 1, 2025'', which could theoretically be delayed.

In light of such shortage of regulations, and building upon the discourse in \citet{policy_probes}'s about using worker platform data to inform policy language, data-sharing tools should help workers and policy experts collaborate on drafting and pushing policies that hold platforms accountable for safer and more just labor standards --- such regulations can provide further resources and transparency for workers experiencing unfair deactivation and suspension, caused by factors such as client accusations. 
Towards understanding workers experiences with safety, P2 suggested surveying workers to learn about the incidents they face and resources platforms provide to protect them. 
In addition to gathering survey and experiential data from workers, we also recommend for worker datasharing tools to go further and build \textbf{a communication channel between workers and policy experts} --- a shared space where both groups can come together to craft and evaluate language being proposed for establishing safety standards or related policy.

%%%%%i took away sharing of resources???
%sharing of resources? 
%\textbf{\anedit{Highlighting Policy Domain Expertise About Relevant Social Services or Programs.}} \anedit{Data-sharing tools can also be designed to let \textit{policy domain experts} share with workers their knowledge about different assistance programs. In interviews, we learned about programs and funding opportunities geared towards aiding workers, many of which align with the initiatives workers and policy domain experts care about, e.g., \textit{Discrimination \& Equity}, \textit{Fair Pay}.} These programs assist citizens seeking employment, guide workers toward higher earning employment (a.k.a. self-sufficiency), or generally provide resources such as working/socialization spaces for workers. \anedit{Often though, information about these programs' existence and support for how to navigate them (e.g., determining eligibility, receiving assistance) is dispersed or confusing. In response, a data-sharing tool can be a centralized place for policy domain experts to raise awareness of things like} training programs, support tools, and resources (e.g., social welfare aid, remediation tactics, affordable health insurance, etc.) \anedit{This may be implemented} through a multi-stakeholder forum, or even a worker-facing resource list. \anedit{This capability also maps to desires that worker participants expressed around the initiative \textit{Experience Sharing}---trainings for how to navigate gig work and manage finances. 

}
%both pointed out power asymmetries in relation to platforms and clients (policymakers--do they know resources? concerns about being in someone's home, different ways safety arises)what are platforms doing to protect them // workers (safety related to discrmination, metrics specifically being a power that clients and platforms have;; inability to say no --> policymakers have a desire to set safety standards for workers.. data for transparency around safety standards and setting new standards . workers--help them crowdsource information experiential or analytical to inform decision-making because they still have to work.

%In fact, a higher awareness about what programs exist and how policy can enact change might empower workers towards collective action 

\subsubsection{\anedit{Balancing Stakeholder-Specific Concerns}}
%Centering Workers’ Voices When Balancing Diverging Stakeholders’ Goals
\anedit{Despite alignments on several prioritized issues, the two stakeholder groups each raised a divergent concern not emphasized by the other.
% Different stakeholders are likely to have distinct priorities, so it was unsurprising that policy experts and workers each raised a concern not emphasized by the other. 
Policy experts wished to understand gig workers' experiences with stress from work and how platforms may exacerbate this (\textit{Stress}). Workers wanted practical assistance to hone their work strategies (\textit{Experience Sharing}). We describe considerations when addressing non-overlapping interests of stakeholders by expanding on these two cases.}
%%%\anedit{Tools aim should be to build understanding, awareness, and common language.}

%Policy stakeholders wanted to understand gig worker stress and how platforms may exacerbate this (\textit{Stress}). Workers wanted practical assistance to hone their work strategies (\textit{Experience Sharing}). 

\paragraph{\anedit{\textbf{\textit{Probing at Latent Worker Needs Raised by Non-Worker Stakeholders.}}}}
\add{While workers themselves may not have the bandwidth to self-assess the impacts of their labor on stress levels, perspectives from policy experts (in addition to existing bodies of work \cite{gkI5, bSah}) helped elucidate the potential need to increase awareness and measurement of such higher-level factors.
% \anedit{First, we underscore that worker-participants not focusing on a certain issue---here, \textit{Work Stress}---does not mean it is not a concern to workers at large. 
%
% However, in the case of diverging motivations, whose goals should be prioritized? Throughout our study, we have been guided by centering gig workers as the domain experts. We maintain that a data-sharing system should actively and directly elevate the goals that workers specifically raise.
%
We maintain that a data-sharing system should actively and directly elevate the goals that workers specifically raise, but this priority on stress exemplifies a case where a datasharing system can leverage insights of stakeholder groups that support workers to help them explore, uncover and become aware of their own latent needs and desires.
% This \textit{may} suggest the presence of latent needs that a data-sharing system can help workers and their supporting stakeholders uncover.
% In particular, system designers may consider ways of 
% \textit{indirectly} raising worker awareness about additional initiatives they may not have considered. 
One feature to support such explorations (while minimizing intrusions to workers) is a \textbf{data collection request form} that 
would require experts to give explanations on what initiatives their requested data is intended to support when they ask for access. This would act as a mechanism that serve the dual purposes of allowing policy domain experts access to data for worker-centered policy while giving workers agency to deliberate on initiatives that they likely care about but are previously unaware of.%The system should also empower workers to take appropriate actions for new initiatives they want to support or are concerned about.
} %\anedit{Forms can be designed to align policy efforts with worker priorities, such as highlighting the need for physical safety and well-being protections (\ref{safety})}.

\paragraph{\anedit{\textbf{Spotlighting Worker Experiences \& Strategies.}}}
\anedit{Workers (understandably) expressed strong desires for practical assistance (e.g., work strategies), but these do not always align with big picture policy initiatives. 
% Earlier, we suggested creating data tools that support education around \textit{Fair Pay} and can offer immediate uses for workers and policy experts. Similarly, 
However, it is still of foremost importance to design for worker-specified goals regardless of whether they overlap with priorities of other stakeholders, since (1) they are the end-users and impacted population of datasharing tools and (2) meeting such priorities offer workers incentivization for engaging in datasharing. 
 % \textit{Experience Sharing}.
%. ---learning about other \textit{Workers' Experiences}---  workers data for practical assistance 
%We advocate for  is important that a data-sharing system offers workers immediate, individual benefits beyond the potential of collective data insights and policy action---both of which will require time to garner enough data for meaningful change.

In this study, workers expressed yearnings for \textit{Experience Sharing}, emphasizing the need and value of qualitative information---anecdotes about clients, psychological well-being support, and answers to frequently asked questions. %\anedit{We offer considerations for incorporating qualitative data collection and sharing in data-exchange tools.}
This suggests potential value in creating systems with capabilities for sharing experiential data. Integrating and extending existing spaces where workers already engage in experience-sharing (e.g., platform-specific sub-reddits, uberpeople.net) \cite{ERsM, woodcock2019gig} is one alternative that reduces upfront effort, but recent work showed how their loosely-organized nature makes it difficult for others to understand and uptake shared narratives \cite{peersupport}. 
Since several worker-participants already identified typologies of sought content (e.g., advice on setting prices and dealing with difficult clients, emotional support via ``ranting'' posts), we recommend designing \textbf{new mechanisms for sharing experiences to cater to needs specific to gig workers} (e.g., search functions, compartamentalization of topics, scaffolding based on expertise).}
% Specifically, these can include 
% Of course, the trade-off of the latter would be accompanying considerations around aspects like} new moderation structures and rules for organizing FAQs.

\subsection{Reflections of Ongoing Challenges}
\anedit{Despite our suggestions for how researchers can design tools that align efforts of stakeholders supporting worker-centered policy, there remain challenges when creating and implementing data-sharing tools.}
%findings 5.3.1 --> challenges around meeting needs of multiple worker types...I CONNECTED DATA INTEGRITY AND INVISIBLE WORK TO THIS
%findings 5.3.2 --> prviacy and trustworthiness (what are the privacy policies, nervous about location data, what if problematic past clients can see it--expand with client data being sensitive too; we have this
%findings 5.3.3 --> ownership and access of data ; %SKIP findings 5.3.4 --> different people to share to
%\textcolor{blue}{maybe add an intro sentence based on jane's orientation of third findings section. something about how beyond designing features for alignment of multi-stakeholder initiatives for worker well-being and policy, we need to consider important issues for the sustainability and usefulness of such a platform as it relates to data integrity, privacy, legislation \& regulation (?), and limitations of data.}
%Limitations of Data
%--risk of overemphasizing quantitative data b/c it may be the most direct to make sense of
%--will create extra work on workers
\subsubsection{Complications with Data Integrity}
%Data Integrity Complications Due to Heterogeneous Preferences for Data Contribution
\anedit{One interesting challenge surfaced from worker workshops was that workers' heterogeneous preferences for sharing data might lead to issues in data integrity.} Maintaining data integrity is important for ensuring usefulness and representativeness for policymakers (a point raised by W2). Yet, \anedit{we learned about} certain work practices \anedit{and data contribution methods that} can degrade data integrity. For example, one petsitter workshop revealed how workers \textit{want to} take clients off the app, but such actions would lead to higher manual effort from workers to collect and input work information. The overheads of manual data entry may \replace{lead to workers being less likely to contribute}{dissuade workers from contributing}, posing challenges to the representativeness of the data. 
Personal informatics research characterized this challenge in data tracking/entry as ``lapsing'' \cite{epstein2015lived} whereby upkeep often ``de-motivates'' users from logging data or results in reduced granularity of data logged \cite{epstein2015lived, li2010stage, choe2014understanding}. 
The costs of manual entry can limit data collection for workers who lack adeptness with technology or time to contribute. W3 discussed how most of her off-app clients are ``financially strapped'' or ``tech savvy''---missing out on worker data associated with these situations can misrepresent or under-represent the needs of lower socioeconomic households, causing second-order effects on actions such as policymakers' decisions for initiatives.
% has explored people's decisions and reasons for selecting, using, and abandoning data collection tools 
%Though we did not hear from freelancers about taking clients off the app, the nature of the work they do and the worker-client relationship they hold could lead to repeated contracting which may be desirable to take off-app, risking data integrity. And while this pattern of long-term or off-app clientele is less typical for drivers, many on-demand gig workers work multiple apps and may feasibly be practicing offline gig work as well. 
Other design considerations for data integrity include ensuring all data sources are captured for multi-app platform workers, developing a methodology for non-automated work data collection that reduces worker burden while maintaining data quality, and validating truthfulness of data.

\subsubsection{Generating New Forms of Invisible Labor}
It is also important to recognize the invisible work that workers perform when contributing non-automatable work data. 
Asking workers who are already data laborers \cite{GUVs} to perform tasks in addition to their jobs (potentially even while they are working) can impose unnecessarily stress---a factor that developers should consider early on to mitigate. 
Relatedly, researchers might also consider ways the design of data-sharing tools can 1) make current invisible labor practices visible to policy domain experts, and 2) alleviate existing invisible work.
To the first point, we reflect that worker participants' responses for data they seek hints at invisible labor they currently perform, such as payment management \cite{toxtli2021quantifying}---e.g., pay transparency variables to understand algorithmically determined wages (\ref{pde_pay})---
and care labor \cite{raval2016standing}---e.g., experiences of others to get tips on handling difficult customers (\ref{h.pplzn44x8y2i}). One idea to address this is centering these forms of data on a policy domain expert dashboard to raise their awareness about \anedit{workers' most critical} initiatives to prioritize. To advance the second point, \anedit{we recommend exploring data-sharing features that} \replace{support}{enable} workers \replace{in}{to} \anedit{intuitively and efficiently navigate, search for, and retrieve qualitative information.} %navigating  repositories of multi-stakeholder sourced worker resources and (\ref{discussion-qualitative}) around supporting the workers' search for qualitative information in an efficient manner.

%%\subsubsection{\anedit{Mitigating an Overreliance on Quantitative Data}}do we need to reflect on this?? could add to the section above about invisible labor 

\subsubsection{Privacy Implications of Data Collection \& Sharing}
%what is the challenge, how might we address it

While many policy expert participants did not express deep concerns about the privacy, security, and ethics practices of a data-sharing system, \anedit{and most} worker-participants conveyed a willingness to partake as long as the system ensured anonymity, this does not preclude risks.
\add{We speculate that this might be a result of participants' lack of familiarity and experience contributing to data donation tools --- it can be difficult to imagine and consider related concerns with data privacy when one doesn't have practical experience engaging with civic tech or data activism.}
D2's desire to revoke data and W3's concern around scope creep remind us to critically consider privacy and ethics criteria when designing worker tools, including how to obtain informed consent and ensure true data ownership. For example, if new data is required as evidence for policy, what consent mechanisms should be in place for new data collections, and how should workers be responsibly informed about risks, benefits, and burdens of the new ask? While identities can be anonymized, workers from places with sparser data contributions could face higher risk of identification if their demographic or location data became exposed. 

\anedit{Considerations about privacy of \textit{client} data also arise. In some cases, client data may be necessary to support initiatives such as \textit{Fair Pay} and \textit{Discrimination \& Inequities}. 
Relevant data can include the payments customers made to platforms to investigate \textit{Fair Pay} or the ratings they gave to workers to investigate instances of \textit{Discrimination}. 
Maintainers and owners of datasharing systems must strike a balance between the goals of (1) collecting and protecting of client data with (2) securely and correctly linking it to corresponding worker data.
% Now only must we ask, how should a system collect and protect client data, but researchers must also wrangle through how to correctly join it with worker data it corresponds to. 
Additionally, sensitive client data could be shared unintentionally within qualitative experiences that workers write about. Such risks can be especially pronounced in caretaking domains, where workers might accidentally include identifiable information about clients when sharing experiences with other workers, especially in cases of safety compromises, such as the case described in \ref{safety}.} \anedit{Additionally, we reflect that our academic conceptualizations of privacy may have limited our ability to surface workers' privacy concerns. Recent work by \citet{kahn2024expanding} points out that whereas ``privacy domain experts'' such as academics often view privacy concerns as violations of privacy laws, consent, monitoring and surveillance. Yet, ``experiential experts'' (in our case, workers) may recognize privacy violations in other ways such as social stigmas around shame and jealousy.  \citet{vashistha2018examining} echoes the importance of exploring workers' concerns around privacy, so that we may situate understandings in relational and experiential contexts that surface localized attitudes and expectations for data-sharing privacy protections. One emerging technique that may mitigate such risks to consumers is the potential of leveraging AI-powered obfuscation techniques to help end-users preserve privacy \cite{monteiro2024manipulate}}
\add{\section{Limitations \& Future Directions}}
We worked with local and national-level policy domain experts in the United States to impact both municipal and national-level changes. Due to overlap in results, we did not separate out results between policymakers of different governing levels --- but this may limit the representatives of insights from each level. We also acknowledge that many states (including the site of this study) restrict the power of authorities to regulate businesses via municipality codes -- limiting regulation of \del{worker-}hostile practices from platforms \cite{collier2018disrupting}\replace{. Thus,}{;} we encourage future work to prioritize efforts \replace{that affect}{affecting} state-level regulations. 
% Furthermore, the catch-all term of ``gig work'' (P9 in \ref{diverse_types}) fails to capture the various ways that workers refer to themselves --- future work should consider ways of bridging this disconnect and focus on meeting specialized needs of each worker type. 
Additionally, our study focuses \replace{addressing challenges in}{ on} the US context, limiting our understanding of the regulatory landscape of other nations. The priorities and initiatives of workers and policymakers from regions such as the Global South \cite{piecework} or EU may diverge from participants of study \cite{Wd16, k6tr}, so we urge and anticipate further investigations specific to non-US regions, and at the international level \cite{WbPL, novitz2020potential, GgXq}.

\add{We would also like to acknowledge the potential of drawbacks of conducting codesign workshops rather than interviews with workers, which can possibly introduce bias towards more dominant voices in each workshop. While we prompted quieter workers to speak up during each session, it is still possible that the time-constrained nature of the workshops can have limited their interactions and self-expressions.} Finally, though participants generated several candidates for owning\del{/governing} a data-sharing system (\ref{ownership}), a lack of consensus leaves room for future investigations.
% \newline
\section{Conclusion}\label{h.2wjlisksix8b}
Our research explored ways of advocating for gig worker needs and policies through the co-design of a data-sharing system. With interviews and workshops with policy domain experts and workers in the US, we revealed prioritized initiatives and gaps in shared understanding that require further alignment. 
Beyond providing design guidelines for data-sharing systems, our results also pose questions of governance and ownership and data integrity, bringing us one stride closer to designing data-sharing alternatives that empower and strength worker rights.
 
\bibliographystyle{ACM-Reference-Format}
\bibliography{references.bib}

% \FloatBarrier
\section{Appendix}
\begin{table*}
  \centering
  \includegraphics[width=.9\textwidth]{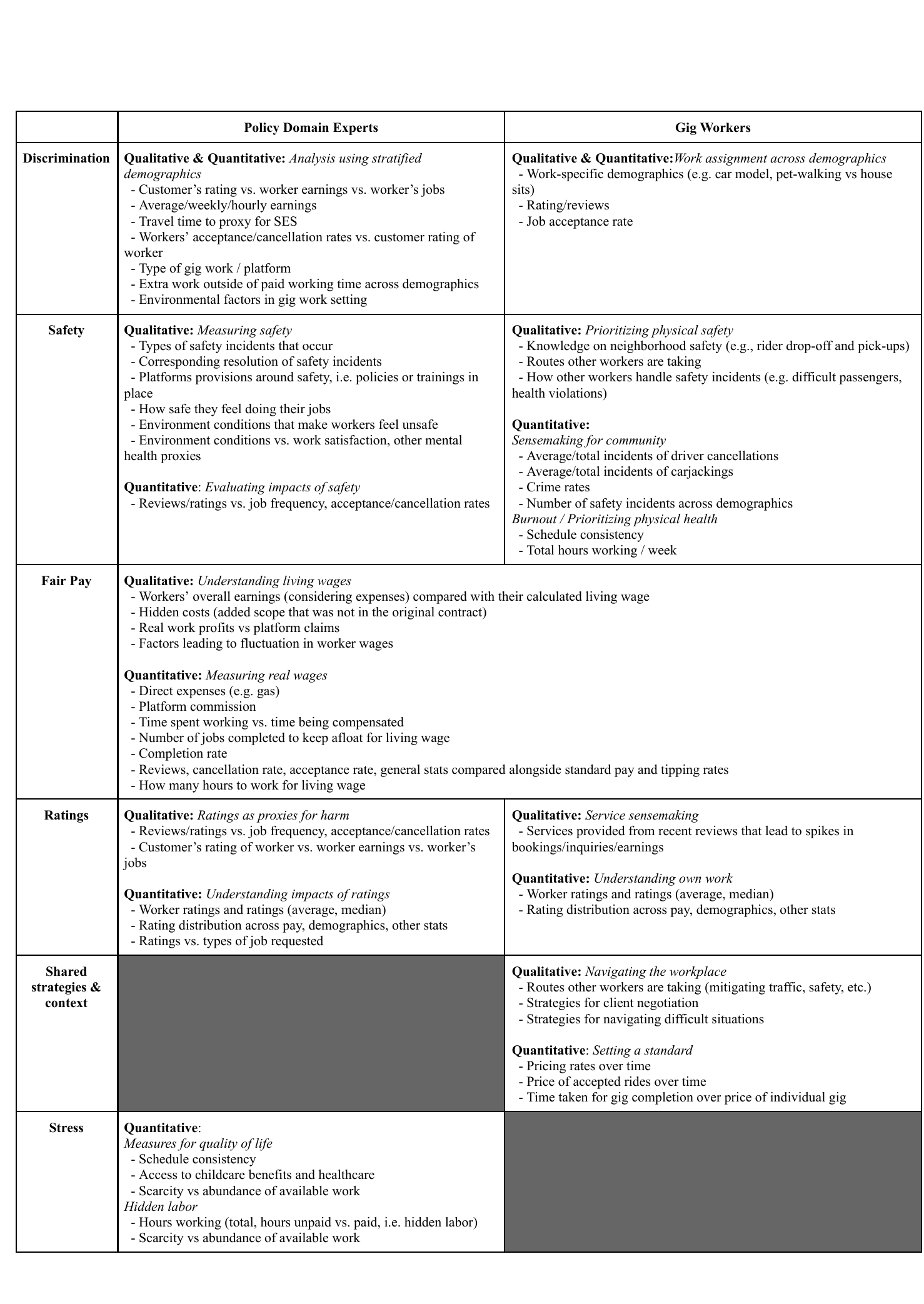}
   \caption{Summary of data needs for initiatives shared across stakeholders}
  \label{tab:data}
\end{table*}
% \FloatBarrier
\end{document}